\documentclass{aa} 

\usepackage{graphicx}
\usepackage{txfonts}
\newcommand{\micron}{\mbox{$\mu$m}}
\newcommand{\tauref}{\mbox{$\tau_z(1~\micron$)}}

\titlerunning{TRUST I. Slab}

\begin{document} 

\title{TRUST I: A 3D externally illuminated slab benchmark for 
dust radiative transfer}

\author{K. D. Gordon \inst{1,2}
  \and M. Baes\inst{2}
  \and S. Bianchi\inst{3}
  \and P. Camps\inst{2}
  \and M. Juvela\inst{4}
  \and R. Kuiper\inst{5,6}
  \and T. Lunttila\inst{8}
  \and K. A. Misselt\inst{10}
  \and G. Natale\inst{9}
  \and T. Robitaille\inst{6,7}
  \and J. Steinacker\inst{11,12,6}
}

  \institute{Space Telescope Science Institute, 3700 San Martin
  Drive, Baltimore, MD, 21218
  \email{kgordon@stsci.edu}
  \and
  Sterrenkundig Observatorium, Universiteit Gent, Krijgslaan 281 S9, 9000 Gent, Belgium
  \and
  INAF-Osservatorio Astrofisico di Arcetri, Largo E. Fermi 5, 50125 Firenze, Italy
  \and
  Department of Physics, P.O.Box 64, FI-00014, University of Helsinki, Finland
  \and
  Institut f\"ur Astronomie und Astrophysik, Universit\"at T\"ubingen, Auf der Morgenstelle 10, D-72076 T\"ubingen, Germany
  \and
  Max-Planck-Institut f\"ur Astronomie, K\"onigstuhl 17, D-69117 Heidelberg, Germany 
  \and
  Freelance Consultant, Headingley Enterprise and Arts Centre, Bennett Road Headingley, Leeds LS6 3HN, United Kingdom
  \and
  Dept.~of Earth and Space Sciences, Chalmers Univ.~of Technology,
  Onsala Space Observatory, SE-439 92 Onsala, Sweden
  \and
  Jeremiah Horrocks Institute, University of Central Lancashire, Preston, PR1 2HE, UK
  \and
  Steward Observatory, University of Arizona, Tucson, AZ 85721
  \and
  Univ. Grenoble Alpes, IPAG,
  F-38000, France
  \and
  CNRS, IPAG, F-38000 Grenoble, France
}

\date{Received by Oct 2016; accepted Apr 2017}
 
  \abstract
   {The radiative transport of photons through arbitrary three-dimensional (3D) structures of dust is a challenging problem due to the anisotropic scattering of dust grains and strong coupling between different spatial regions.
   The radiative transfer problem in 3D is solved using Monte Carlo or Ray Tracing techniques as no full analytic solution exists for the true 3D structures.} 
   {We provide the first 3D dust radiative transfer benchmark composed of a slab of dust with uniform density externally illuminated by a star.  
   This simple 3D benchmark is explicitly formulated to provide tests of the different components of the radiative transfer problem including dust absorption, scattering, and emission.}
   {The details of the external star, the slab itself, and the dust properties are provided.
   This benchmark includes models with a range of dust optical depths fully probing cases that are optically thin at all wavelengths to optically thick at most wavelengths.
   The dust properties adopted are characteristic of the diffuse Milky Way interstellar medium.
   This benchmark includes solutions for the full dust emission including single photon (stochastic) heating as well as two simplifying approximations: One where all grains are considered in equilibrium with the radiation field and one where the emission is from a single effective grain with size-distribution-averaged properties.
   A total of six Monte Carlo codes and one Ray Tracing code provide solutions to this benchmark.}
   {The solution to this benchmark is given as global spectral energy distributions (SEDs) and images at select diagnostic wavelengths from the ultraviolet through the infrared.
   Comparison of the results revealed that the global SEDs are consistent on average to a few percent for all but the scattered stellar flux at very high optical depths.
   The image results are consistent within 10\%, again except for the stellar scattered flux at very high optical depths.
   The lack of agreement between different codes of the scattered flux at high optical depths is quantified for the first time.
   Convergence tests using one of the Monte Carlo codes illustrate the sensitivity of the solutions to various model parameters.}
   {We provide the first 3D dust radiative transfer benchmark and validate the accuracy of this benchmark through comparisons between multiple independent codes and detailed convergence tests.}


   \maketitle
%
%________________________________________________________________

\section{Introduction}

The transport of radiation through dust plays a central role in many astrophysical objects as dust is efficient in absorbing and scattering ultraviolet (UV) through near-infrared photons and re-radiating the absorbed energy in the infrared and submillimeter (submm).  
The interpretation of observations from the UV to the submm is critically linked to an accurate calculation of the radiative transfer (RT) through dust.  
The dust RT problem is  difficult to solve as the time-independent version is six dimensional (space, angle, and wavelength) with strongly anisotropic scattering and non-linear coupling between different spatial regions.
The complexity of the solution is especially evident when the object of interest is intrinsically three-dimensional (3D).  
Solutions using Ray-Tracing and Monte Carlo techniques (and mixtures of the two) are used in modern RT codes to solve this problem.  
\citet{Steinacker13} gives an in-depth review of the 3D RT problem, current solution techniques, and an overview of existing codes, the number of which has grown significantly in the last 15 years.

While it is common to provide benchmark solutions for specific objects to ensure that different codes produce the same answer within some tolerance, there are no existing intrinsically 3D RT benchmarks; existing benchmarks focus on one-dimensional (1D) or two-dimensional (2D) objects  \citep{Ivezic97, Pascucci04, Pinte09}.  
In addition to being 1D or 2D, existing RT benchmarks do not include the full dust radiative transfer solution, using approximations in either dust scattering (e.g., isotropic only) or dust emission (e.g., equilibrium only emission) processes.  
This has motivated a group of 3D dust RT coders to come together and propose a suite of 3D benchmarks that will test the many aspects of the dust RT solution in a range of geometries. 
This suite of benchmarks is named TRUST (benchmarks for the Transport of Radiation through a dUSTy medium)\footnote{http://ipag.osug.fr/RT13/RTTRUST/}.

In general, code benchmarks are motivated by and designed to test for coding errors, ensure accurate calculations, compare how differences between codes impact the results, and test the relative speed of different codes.  
The TRUST effort is focusing on the first three goals as testing the speed of codes is of much lower priority than ensuring that the codes are accurate, error free, and produce consistent results.

Ideally a full analytic solution would be adopted as the target solution for the TRUST benchmarks as this would allow for all benchmarking goals to be achieved.  
Unfortunately, no such analytic solutions exist for the dust radiative transfer equation for any geometries that are intrinsically 3D.  
Thus for 3D dust RT we are left with using a converged solution that has been validated by multiple codes, ideally with different solution techniques.  
This should ensure that the benchmark solution is not affected by coding errors and is likely to be correct.  
This paper presents the results for a geometrically simple, yet still 3D, benchmark.  
This first simple benchmark is an externally illuminated slab of dust and is presented specifically to test the components of dust RT at the basic level. 
Future benchmarks will test specific capabilities of codes in more complex geometries including shells, filaments, shadowed regions, and galaxy disks with spiral structures.

%
%______________________________________________________________

\section{Setup}

\begin{table}
\caption{Slab setup details}
\label{tab_slab_setup}
\begin{tabular}{lc}
\hline\hline
Name & Values \\
\hline
\multicolumn{2}{c}{Dust Geometry} \\
\hline
system size & 10 $\times$ 10 $\times$ 10 pc \\
system coordinates & -5 to 5 pc \\
slab z extent & -2 to -5 pc \\
slab xy extent & -5 to 5 pc \\
\tauref & 0.01, 0.1, 1, 10 \\
\hline
\multicolumn{2}{c}{Blackbody Source} \\
\hline
location (x,y,z) & (0 pc, 0 pc, 4 pc) \\
temperature & 10\,000 K \\
luminosity & 100\,000 L$_\sun$ \\
    & $3.839 \times 10^{38}$~ergs~s$^{-1}$ \\
\hline
\multicolumn{2}{c}{Observer} \\
\hline
distance & 10\,000 pc \\
\hline
\end{tabular}
\end{table}

\begin{figure}
\resizebox{\hsize}{!}{\includegraphics{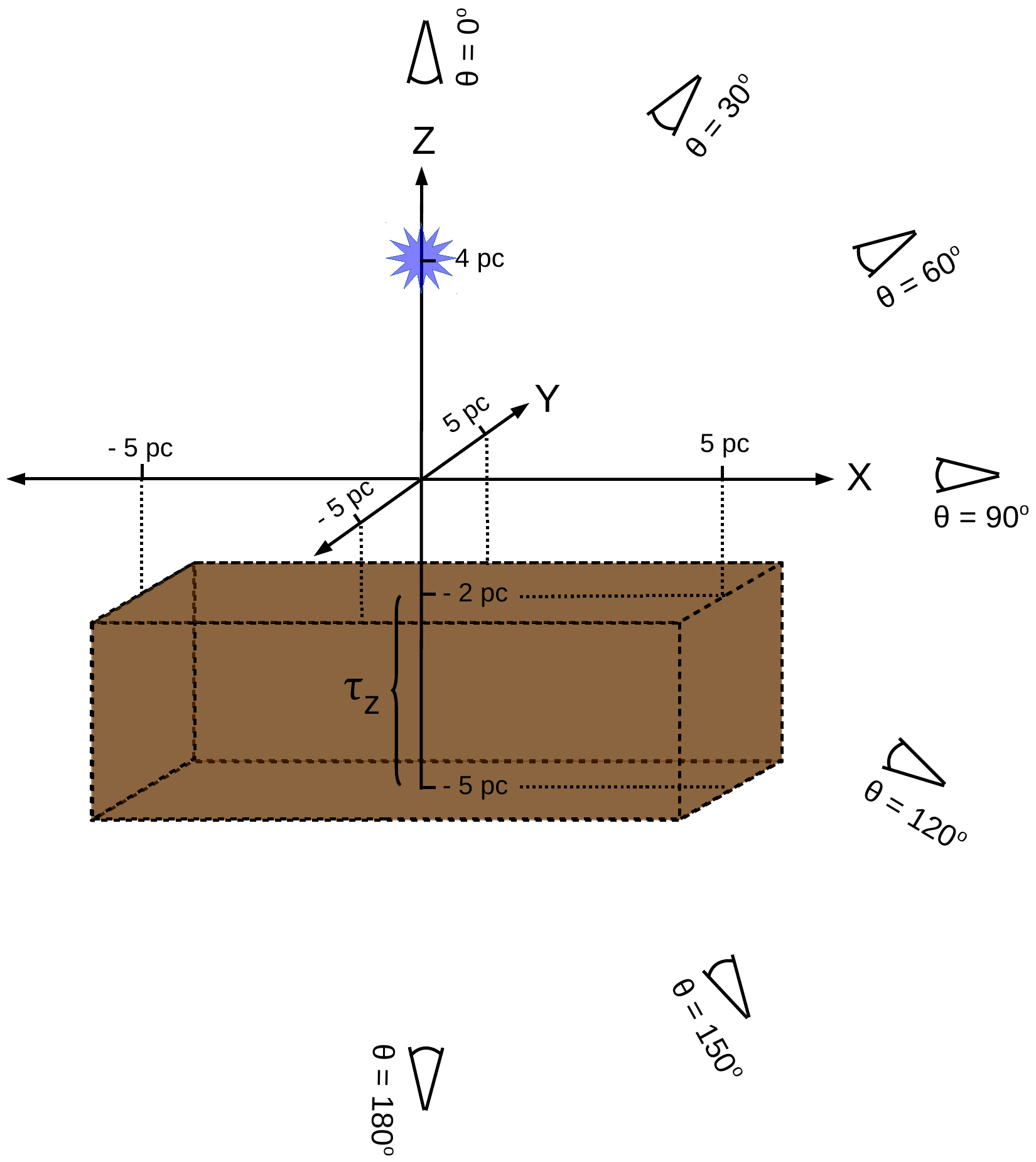}}
\caption{The setup of the slab benchmark is graphically illustrated.  The external views for the output SEDs and images are shown.}
\label{fig_setup_schematic}
\end{figure}

The overall geometry of this benchmark is a rectilinear slab that is externally illuminated by a single blackbody source.  
The values for the slab and point source are given in Table~\ref{tab_slab_setup} and the geometry illustrated in Figure~\ref{fig_setup_schematic}.  
The luminosity of the source is set by integrating the source spectral energy distribution (SED) between 0.09 and 2100~\micron.
All the dust in the system is uniformly distributed through the slab with the rest of the model space being completely empty except for the blackbody source.

\begin{figure}
\resizebox{\hsize}{!}{\includegraphics{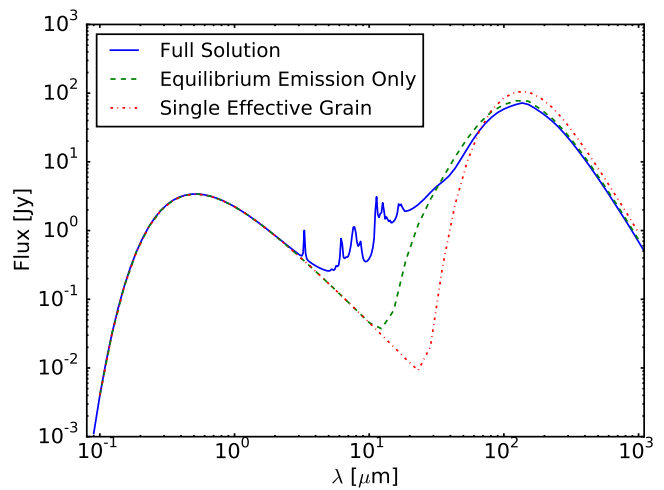}}
\caption{The global SEDs are shown with the three different dust emission choices.
The example shown is for $\tauref = 1$ and $\theta = 90^\circ$.}
\label{fig_example_emission_type}
\end{figure}

The dust grains have the BARE-GR-S properties from \citet{Zubko04} and are discussed in detail by \citet{Camps15b}.  
Given the computational complexity of including the full treatment of dust emission in the RT solution, this benchmark provides results including stochastic heating (sto, the full solution) as well as the equilibrium only heating (equ) and the single effective grain (eff) approximations.  
In the effective grain approximation, the grain properties are integrated over the grain size distribution and summed over the grain components to produce a single grain with effective properties.  
This effective grain is an extreme approximation \citep{Steinacker13} that has the benefit of allowing fast calculation of the dust emission spectrum from the radiation field.  
The equilibrium-only dust emission approximation assumes all grains are in equilibrium with the radiation field, even for those smaller grains that, physically, should be stochastically heated. 
The differences in the global SED from a model with the three different methods for calculating the emission from the dust are illustrated in Fig.~\ref{fig_example_emission_type}.
This figure clearly illustrates the importance of including the full solution to achieve accurate mid-IR and, to a lesser extent, far-IR results for cases like those in this benchmark.
Data files that contain the full grain properties are available online\footnote{http://www.shg.ugent.be} \citep{Camps15b}.

The density of the slab is set by the optical depth along the z-direction, defined as \tauref.
The \tauref\ values chosen here (Table~\ref{tab_slab_setup}) provide a full sampling of optical depths from optically thin to thick.
For $\tauref=0.01$, the model is optically thin at all wavelengths with a maximum of $\tau_z(0.09~\micron) \sim 0.18$.  
The next $\tauref$ value of 0.1 is optically thick in the UV with $\tau_z(0.09~\micron) \sim 1.8$. 
For $\tauref = 1$, the slab is optically thick to all UV and optical photons.
Finally, the $\tauref = 10$ case is optically thick for all $\lambda \lesssim 4~\micron$ and very optically thick in the UV $\tau_z > 100$.

Each RT code generates global SEDs and spatially resolved, multi-wavelength images.
These outputs from each code are compared against all the other codes to estimate the fidelity of the RT solution for each configuration. 
We choose to use outputs for comparison as these quantities are what is generally compared with observations and the internal representation of quantities in RT codes (e.g., radiation field density) are often stored on quite different spatial grids.
Previous benchmarks have also compared the internal dust temperature structure.  We do not as only for the eff grain approximation is a single dust temperature defined.  For the other two cases (equ and sto), there are a range of dust temperatures in each grid cell given the range of grain sizes and compositions.  
Global SEDs are computed for the total as well as decomposed into the different RT components (direct stellar, scattered stellar, direct dust emission, and scattered dust emission).

\begin{figure}
\resizebox{\hsize}{!}{\includegraphics{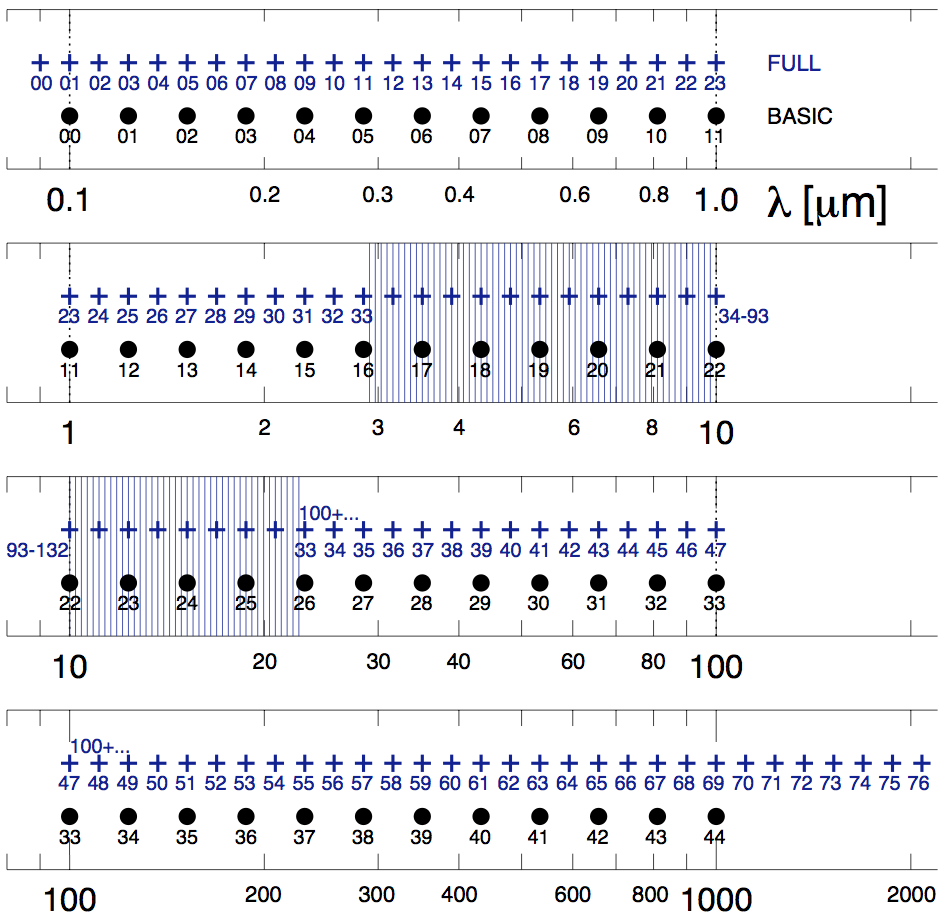}}
\caption{The BASIC and FULL wavelength grids are shown.  
The BASIC grid is used for the models using the effective grain and equilibrium-only emission approximations.
The FULL grid is used for the models computing the full dust emission including stochastically heated grains. Vertical blue lines 
between $\sim$3-23\micron\ represent the dense sampling of wavelength points to resolve the PAH emission.}
\label{fig_wavegrid}
\end{figure}

\begin{table}
\caption{Physical constants}
\label{tab_constants}
\begin{tabular}{ll}
\hline\hline
Constant & Description \\ \hline
$c = 2.99792458 \times 10^8$ m s$^{-1}$ & speed of light \\
$h = 6.62606957 \times 10^{-34}$ m$^2$ kg s$^{-1}$ & Planck constant\\
$k = 1.3806488 \times 10^{-23}$ m$^{^2}$ kg s$^{-2}$ K$^{-1}$ & Boltzmann constant \\
\hline
\end{tabular}
\end{table}

The SEDs are output in units of Jy and images are output in units of MJy/sr. 
SEDs and images are generated at seven viewing angles (0\degr, 30\degr, 60\degr, 90\degr, 120\degr, 150\degr, and 180\degr; see Figure~\ref{fig_setup_schematic}), probing the full range of scattering angles from back-scattering ($\theta = 0\degr$) to forward scattering ($\theta = 180\degr$).  
At lower optical depths ($\tauref = 0.01, 0.1, 1.0$), the resolution of the images is 300$\times$300~pixels while at $\tauref = 10$, the image resolution is 600$\times$600~pixels.
The resolution is set by the need to resolve RT effects in the front surface of the dust slab in the infrared.  
In all cases, the physical size covered by the images is 15$\times$15~pc to account for all possible rotations of the system.  
The two wavelength grids used are shown in Fig.~\ref{fig_wavegrid} and can be obtained from the TRUST website\footnote{http://ipag.osug.fr/RT13/RTTRUST/}.
The BASIC wavelength grid is used for the effective grain and equilibrium only emission approximations as the equilibrium grain temperatures only depend on the the total absorbed energy.
The FULL wavelength grid is used for the full emission solution as the calculation of the stochastically heated grains requires a finer resolution sampling to calculate the temperature probability distribution.
The FULL wavelength solution includes a very fine spacing in the mid-IR to resolve the aromatic/PAH emission features.
Finally, we give the adopted values of relevant physical constants in Table~\ref{tab_constants} as their exact values will change the output SEDs and images.

\subsection{Example outputs\label{sec:ex_outputs}}

\begin{figure*}
\resizebox{\hsize}{!}{\includegraphics{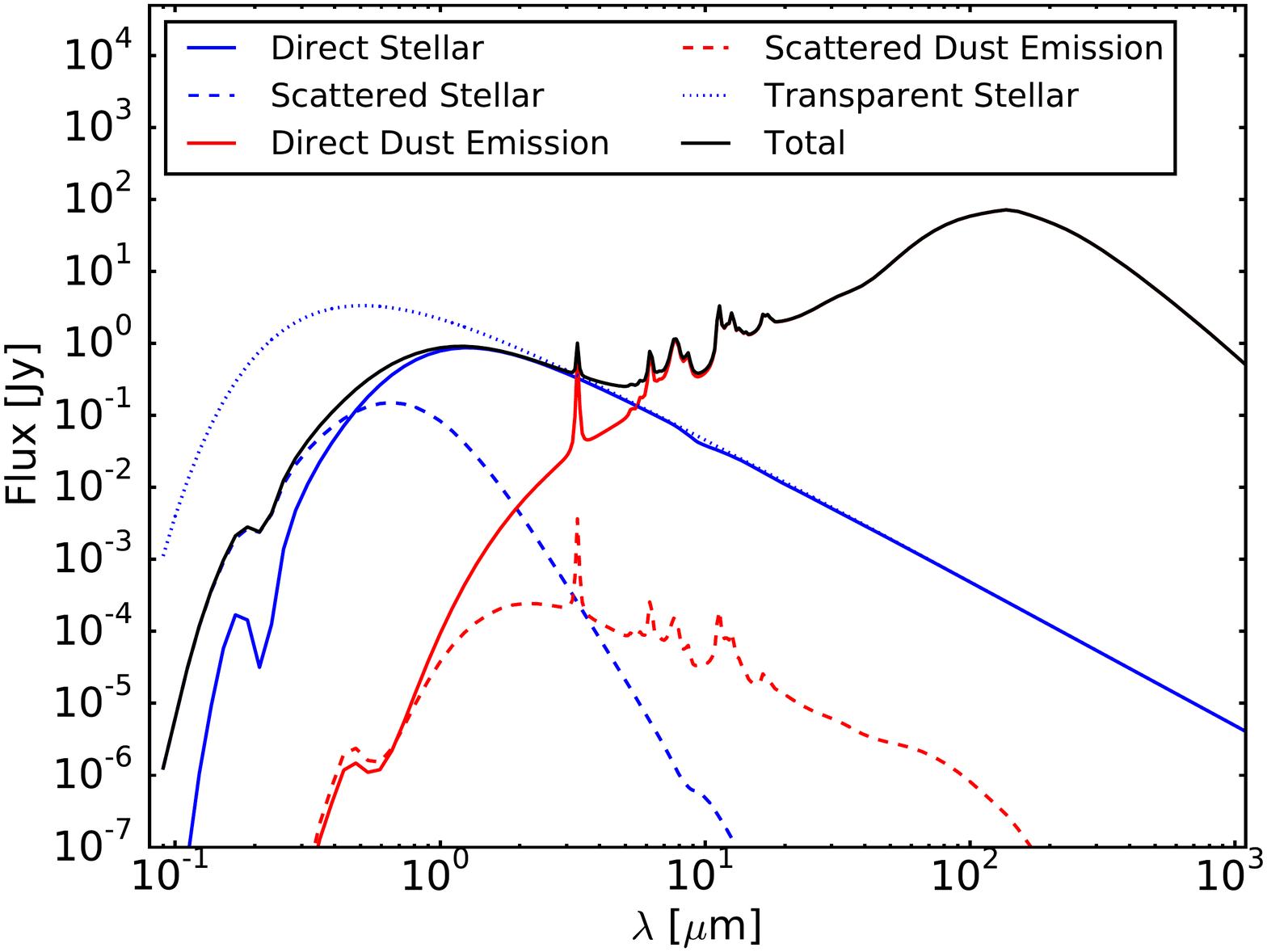}
    \includegraphics{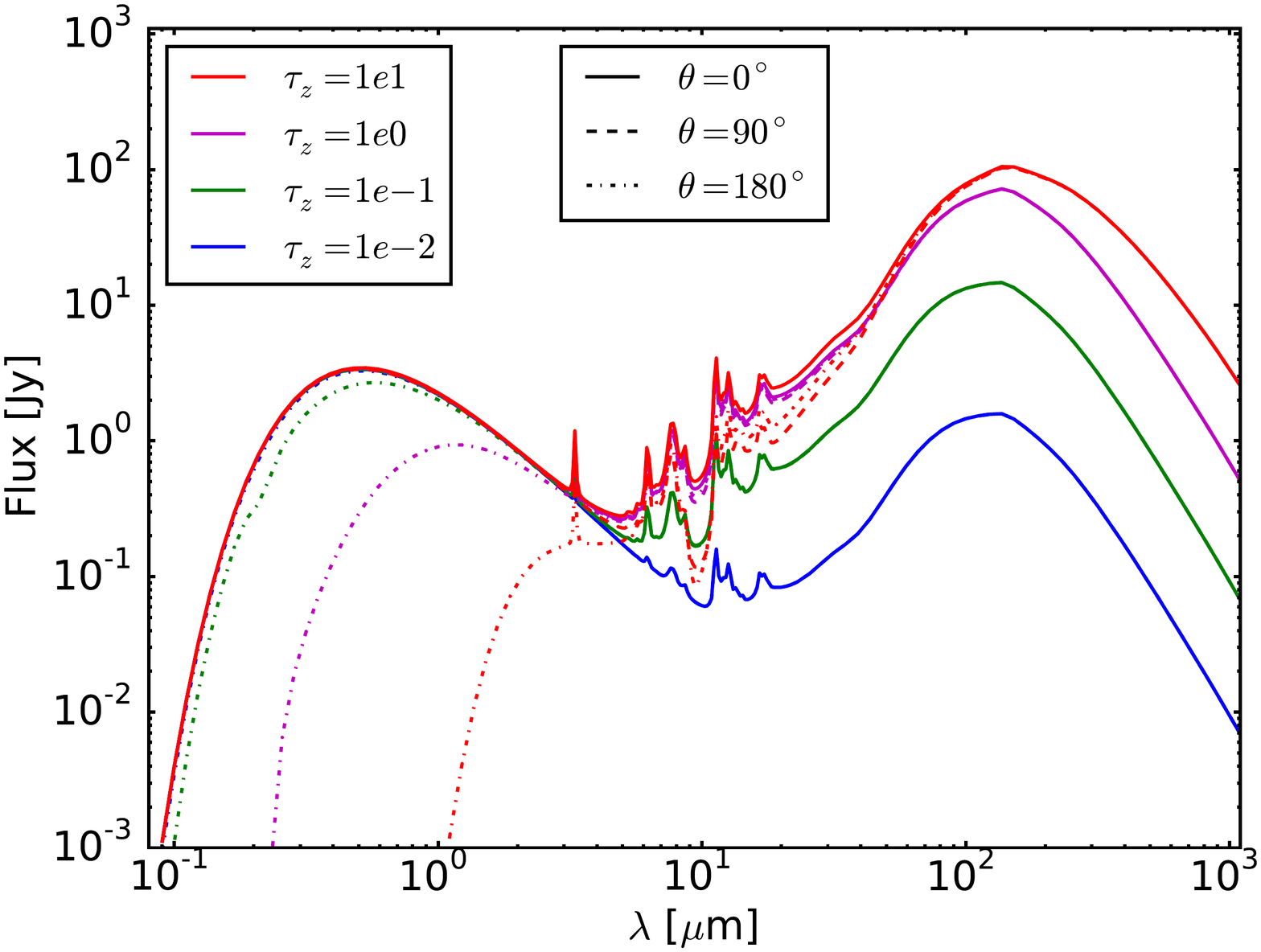}}
\caption{Examples SEDs are shown from models run with the full dust emission solution.  
On the left, the total and components of the global SED are shown for the $\tauref = 1$ and $\theta = 150^\circ$ case.
On the right, the total SEDs are shown for all \tauref\ values and $\theta$ values of $0^\circ$, $90^\circ$, and $180^\circ$.}
\label{fig_example_sed_components}
\end{figure*}

Fig.~\ref{fig_example_sed_components} gives an example of the global SED outputs.
On the left, the total SED (Total) is shown along with the different components that comprise the total.
Decomposing the total SED into components is important to test the differences in the different parts of the dust RT solution between models. 
The components include the stellar flux attenuated by any line-of-sight dust (Direct Stellar), the stellar flux scattered by the dust (Scattered Stellar), the thermal emission from the dust (Direct Dust Emission), and the scattered dust emission (Scattered Dust Emission).  
In addition, the stellar flux from the dust-free model (Transparent Stellar) is also shown as it is diagnostic of how each code treats the input stellar photons.
The particular example shown illustrates the importance of the dust scattered stellar component in the ultraviolet and optical.
The right panel shows the total SEDs covering the full \tauref\ and $\theta$ range.
The $\theta = 180^\circ$ SEDs at shorter wavelengths clearly illustrate the impact of observing the star through the dust slab.
The impact of dust self-absorption can be seen at $\tauref = 10$ most easily from the increasing depth of the silicate absorption feature at $\sim$$10~\micron$ with increasing viewing angle $\theta$.

\begin{figure*}
\resizebox{\hsize}{!}{\includegraphics{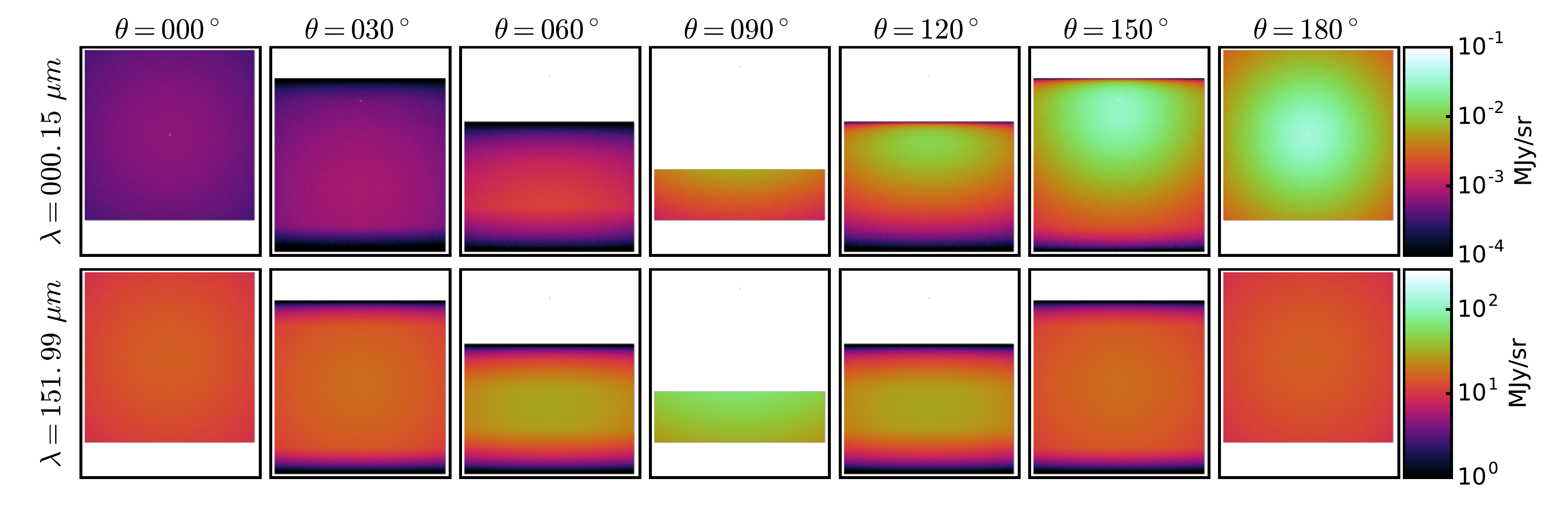}}
\caption{The output images from the SKIRT model are shown for the $\tauref = 0.1$ case with the full dust emission solution for two representative wavelengths.}
\label{fig_example_images}
\end{figure*}

Examples of the output images are shown in Fig.~\ref{fig_example_images} for the full range of viewing angles $\theta$.  
The output SED at $\lambda = 0.15~\micron$ is dominated by scattering and the corresponding images illustrate the strongly forward scattering nature of dust grains with the brightness of the scattered light increasing from $\theta = 0^\circ$ (back scattering) to $\theta = 180^\circ$ (forward scattering).
The $\lambda = 151.99~\micron$ images show that the infrared emission has a different dependence on viewing angle compared to the  $\lambda = 0.15~\micron$ images, symmetrically peaking at $\theta = 90^\circ$, reflecting the isotropic nature of the dust emission. 

All the model outputs for all the codes are available online\footnote{https://doi.org/10.5281/zenodo.163632}.

\section{Models}

There are seven 3D dust RT codes that participated in this benchmark.  
Six of them are based on Monte Carlo techniques (CRT, DIRTY, Hyperion, SKIRT, TRADING, and SOC) and one is based on Ray-Tracing (DART-Ray).  
Here we provide a very brief description of each code.
The reader is encouraged to read the references provided for the details of how each code implements the solution to the RT problem.

\subsection{CRT}

CRT is a 3D Monte Carlo radiative transfer code \citep{Juvela2003, Lunttila2012}.
It uses nested Cartesian grids for representing the dust distribution.
The program can use several methods for accelerating calculations, for example packet weighting \citep{Juvela2005}, polychromatic photon packets \citep{Jonsson2006}, and subgrid iteration \citep{Lunttila2012}.
CRT includes a high-performance library for computing emission from arbitrary dust models, including stochastically heated grains \citep[see][]{Camps15b}; however, it also allows using an external program for dust emission calculations \citep[see, e.g.,][]{Ysard2011}.
The main application of CRT has been studies of molecular clouds and Galactic star formation.
In particular, it has often been used for creating synthetic scattered light and dust thermal continuum observations of cloud models from magnetohydrodynamic simulations to help quantify the uncertainty in observationally derived cloud properties such as column density and core mass distribution \citep[e.g.,][]{Juvela2006, Malinen2011}.
Other applications have included stability analysis of non-isothermal Bonnor-Ebert spheres \citep{Sipila2011} and galaxy mergers \citep{Karl2015}.

\subsection{DART-Ray}

DART-Ray is a purely ray-tracing 3D dust radiative transfer code. 
Its core algorithm has been presented by \citep{Natale14} and further developed by Natale et al. (2017, in preparation). 
This algorithm reduces the amount of calculations that would be required in a brute-force ray-tracing approach by limiting the propagation of rays from each radiation source throughout the RT model within the source influence volume. 
The latter is the volume around a radiation source within which the source contributes significantly to the radiation field energy density. 
This code utilises adaptive Cartesian grids to define the distributions of stars and dust and an optimization technique to set the angular density of rays cast from each source. 
The dust emission can be calculated both for dust in equilibrium with the radiation field and dust that is stochastically heated \citep{Natale15}. 
The current version of DART-Ray does not include dust self-heating and dust emission scattering. 
In the context of this benchmark, the results are expected to be accurate in the infrared only for the cases where the dust emission is optically thin. 
We note that, in contrast with the other participating codes, the maximum number of scattering iterations is not set by the user in DART-Ray. 
Instead, scattering iterations are stopped when the remaining scattered radiation luminosity to be processed is less than 1\% of the initial scattered radiation luminosity.    

\subsection{DIRTY}

The DIRTY radiative transfer code is a 3D Monte Carlo radiative transfer code \citep{Gordon01, Misselt01}.  
It can handle arbitrary geometries efficiently using nested Cartesian grids.
It includes a full dust grain model description allowing for arbitrary dust grain models to be used.
The dust emission calculation includes the full solution including both equilibrium and non-equilibrium (stochastically heated) emission.
This code has been used to study the general behavior of radiative transfer through clumpy dust \citep{Witt96}, Milky Way reflection nebulae \citep{Gordon94, Calzetti95, Lewis09}, regions of galaxies \citep{Witt00, Misselt01}, starburst galaxies locally \citep{Gordon97, Gordon00, Law11} and at high redshift \citep{Vijh03}, and disk galaxies \citep{Pierini04}.
For this benchmark, the spacing in the z direction was log and linear in the x and y directions.
Motivated by this benchmark, the composite biasing technique \citep{Baes16} was added to better sample scattering at high optical depths and the dust emission from radiation fields with strong spatial variations.

\subsection{Hyperion}

Hyperion \citep{Robitaille11} is an open-source\footnote{http://www.hyperion-rt.org/} dust continuum 3D Monte Carlo radiative transfer code. 
It is designed to be modular, and can be used to compute radiative transfer for arbitrary 3D geometries and has been applied to, for example, analytical models of forming stars \citep{Robitaille14,Koepferl15,Johnston15}, galaxy formation simulations \citep{Narayanan15}, and large-scale Galactic emission \citep{Robitaille12}. 
A number of grid geometries are supported, including Cartesian, spherical and cylindrical polar, nested Cartesian, octree, and Voronoi grids. 
Grid cells are never required to be regularly spaced, so for the models presented here, the cells in the $z$ direction are spaced logarithmically (with the first cell below the surface of the slab located at $-2.001$). 
Multi-process parallelization is implemented using MPI and used for this benchmark. 
Hyperion supports computing the radiative transfer for one or more dust populations, but does not include full support for stochastic heating (instead, computing models with small grains and PAHs can be done using precomputed templates, as done in \citealt{Robitaille12}). 
Since the original implementation of the code presented in \citet{Robitaille11}, forced first scattering using the \citet{Baes16} algorithm has been added (and the results here assume $\xi=0.2$). 
In addition, the process for monochromatic radiative transfer (described in \S2.6.4 of \citealt{Robitaille11}) has now changed - when computing the contribution to the scattered light emission, photon packets are scattered and lose energy until their energy goes below a certain threshold (set to $10^{-120}$ for this paper), as opposed to randomly sampling whether to scatter or absorb at each interaction (which caused many scattering scenarios to be under-sampled and therefore required large numbers of photons in order to attain a good signal-to-noise in certain situations).

\subsection{SKIRT}

SKIRT is a public\footnote{http://www.skirt.ugent.be} multi-purpose 3D Monte Carlo dust radiative transfer code \citep{Baes11, Camps15a} for simulating the effect of dust on radiation in astrophysical systems.
It offers full treatment of absorption and multiple anisotropic scattering by the dust, computes the temperature distribution of the dust and the thermal dust re-emission self-consistently, and supports stochastic heating of dust grains \citep{Camps15b}.
The code handles multiple dust mixtures and arbitrary 3D geometries for radiation sources and dust populations, including grid- or particle-based representations generated by hydrodynamical simulations.
The dust density distribution is discretized using one of the built-in dust grids, including state-of-the art octree \citep{Saftly13}, $k$-d tree \citep{Saftly14}, and Voronoi \citep{Camps13} grids.
The wide range of built-in components can be configured to construct complex models in a parameter file or through a user-friendly interface \citep{Camps15a, Baes15}.
SKIRT implements hybrid parallelization, allowing an arbitrary mix of multiple threads and processes possibly across multiple computing nodes (Verstocken et al., in preparation).
While SKIRT is predominantly used to study dusty galaxies \citep[e.g.,][]{Baes02, Baes10, Baes11, DeLooze12, DeGeyter14,Saftly15}, it has also been applied to active galactic nuclei \citep{Stalevski12}, molecular clouds \citep{Hendrix15}, and stellar systems \citep{Deschamps15}.

\subsection{TRADING}

TRADING \citep{Bianchi08} is a 3D Monte Carlo radiative transfer code originally designed to study the effects of clumping in simulations of dust extinction and emission in spiral galaxies.
Developed from an earlier regular-grid, thermal-emission-only code \citep{Bianchi96,Bianchi00}, TRADING includes an octree grid for the dust distribution, stochastic heating, and dust self-absorption.
It neglects scattering of dust-emitted radiation.
It has been used to model dust extinction \citep{Bianchi07} and emission \citep{Bianchi11,Holwerda12} in edge-on galaxies, and to study the dust heating in the host galaxy of high-z Quasi-Stellar Objects \citep{Schneider15}.
A few adaptations were made to TRADING for this benchmark: while each photon packet wavelength is drawn from a (tabulated) probability distribution function (PDF), at odds with other RT codes, the weight of the photons was adjusted so that a similar number of packets falls in each bin of a logarithm-spaced grid (i.e.,\ a weight $1/\lambda$ was used).
Forced scattering was used for the first scattering event and along all paths with $\tau > 5$, applying the composite biasing scheme described by \citet{Baes16} for half the events along these paths. 
The models presented here were all run using an octree grid of $\approx 7.5\times 10^{5}$ cells, having a higher resolution in the part of the slab facing the source. 

\subsection{SOC}

SOC is a new Monte Carlo radiative transfer code that is parallelized using OpenCL libraries (Juvela et al., in preparation). 
The models can be defined on regular Cartesian, modified Cartesian, or octree grids. 
In this paper, calculations employ modified Cartesian grids that are defined by cell boundaries located at arbitrary positions along the three main axes. 
In dust-scattering calculations, SOC uses photon packages that each contain photons from four consecutive frequency bins. 
The differences between the optical depths and scattering functions at those frequencies are compensated by weighting \citep[see][]{Juvela2005, Jonsson2006}. 
Calculations for dust emission proceed one frequency at a time. 
The dust temperature distributions and the dust emission (per cell) of stochastically heated grains are calculated with an external program that is common with CRT. 
For dust models assuming an equilibrium temperature, calculations are done within SOC itself.

\section{Output comparisons}

The comparison between the results from the different codes is based on measuring the average absolute deviation from the median result from all the codes. Explicitly, for each code ($j$) we calculate
\begin{equation}
\bar{\Delta_j} = \frac{1}{n} \sum_i{|(x_{ij} - \mu_i)|}
,\end{equation}
where $n$ is the number of wavelength/spatial points, $x_i$ is the $i$th wavelength/spatial point for the global SED/image, and 
\begin{equation}
\mu_i = \mbox{med}_j(x_{ij}).
\end{equation}
We use the median for comparison as there is no analytic solution and use of a median is more robust to outliers.
Given this metric, the deviation of one code from the reference being smaller than the deviation of other codes is not significant per se; it simply means that it is nearer the median. 
The goal of these quantitative comparisons is to determine the overall agreement between codes and to determine whether or not the differences between code results are random or systematic.

\begin{table*}
\caption{Code parameter values}
\label{tab_model_params}
\begin{tabular}{llccccccc}
\hline\hline
Name & Description &  CRT & DART-ray & DIRTY & Hyperion & SKIRT & TRADING & SOC \\
\hline
$N$ & \# of photons/rays per wavelength & $10^9$ & $10^9$--$10^{10}$ & $10^8$ & $10^9$ & $10^8$ & $2\times 10^8$ & $5\times 10^8$\\ 
$n_{xy}$ & \# of bins in $x$ or $y$ & 100 & 90 -- 992 & 100 & 100 & 100 & 40 -- 160 & 150 \\
$n_z$ & \# of bins in $z$ & 150 -- 300 & 29 -- 31 & 100 & 200 & 100 & 144 & 150 \\ 
$m_\mathrm{scat}$ & max \# of scatterings & $\sim$60 & 8 & 500 & $\sim$270 & $\sim$250 & $\sim$390 & 20 \\
$m_\mathrm{iter}$ & max \# of dust heating iterations & 5 & 0 & 4 & 5 & 50 & 4 & 4 \\ \hline
\end{tabular}
\end{table*}

As certain parameters are particularly important for the precision of the results, as revealed by the convergence tests (\S\ref{sec_convergence}), we give the values for such parameters for all the codes in Table~\ref{tab_model_params}.
The number of rays/photons per wavelength ($N$) is only roughly comparable between different codes as, for example, the emission from the stellar source may be biased towards the slab or not.
In addition, rays or photons may be split multiplying the impact of a single ray or photon.
The value of $m_\mathrm{scat}$ is also only roughly comparable between codes as the scattering may be done by forcing no scattering, forcing the first scattering, or forcing all scatterings.
The maximum number of heating iterations $m_\mathrm{iter}$ controls whether or not dust self-heating is taken into account and how many iterations are allowed to achieve a specified convergence level.
If $m_\mathrm{iter} = 0$, then dust self-heating is not accounted for and the dust emission is only determined
by the absorbed stellar photons.
While these parameters are only roughly comparable between codes, they do provide a qualitative comparison between codes of the precision of different radiative transfer solution components.

While most of the codes include scattering of the dust emission, DART-ray and TRADING do not.
This can affect the accuracy of the infrared images from these codes, but it is unlikely to affect the global SEDs as the total scattered dust emission is negligible compared to the total direct dust emission.

The adopted distance was not used fully in creating the images by all the different codes.
Some codes took into account the projection effects due to the finite distance while others assumed an infinite distance resulting in no projection affects.
The resulting differences are small in the images, except for the pixels sampling the edge of the slab and are noticeable for a
viewing angle of $\theta = 90^\circ$. These edge pixels were not used for the image comparisons to avoid injecting differences that are purely geometrical into the comparisons.

\subsection{Example comparisons}

\begin{figure*}
\resizebox{\hsize}{!}{\includegraphics{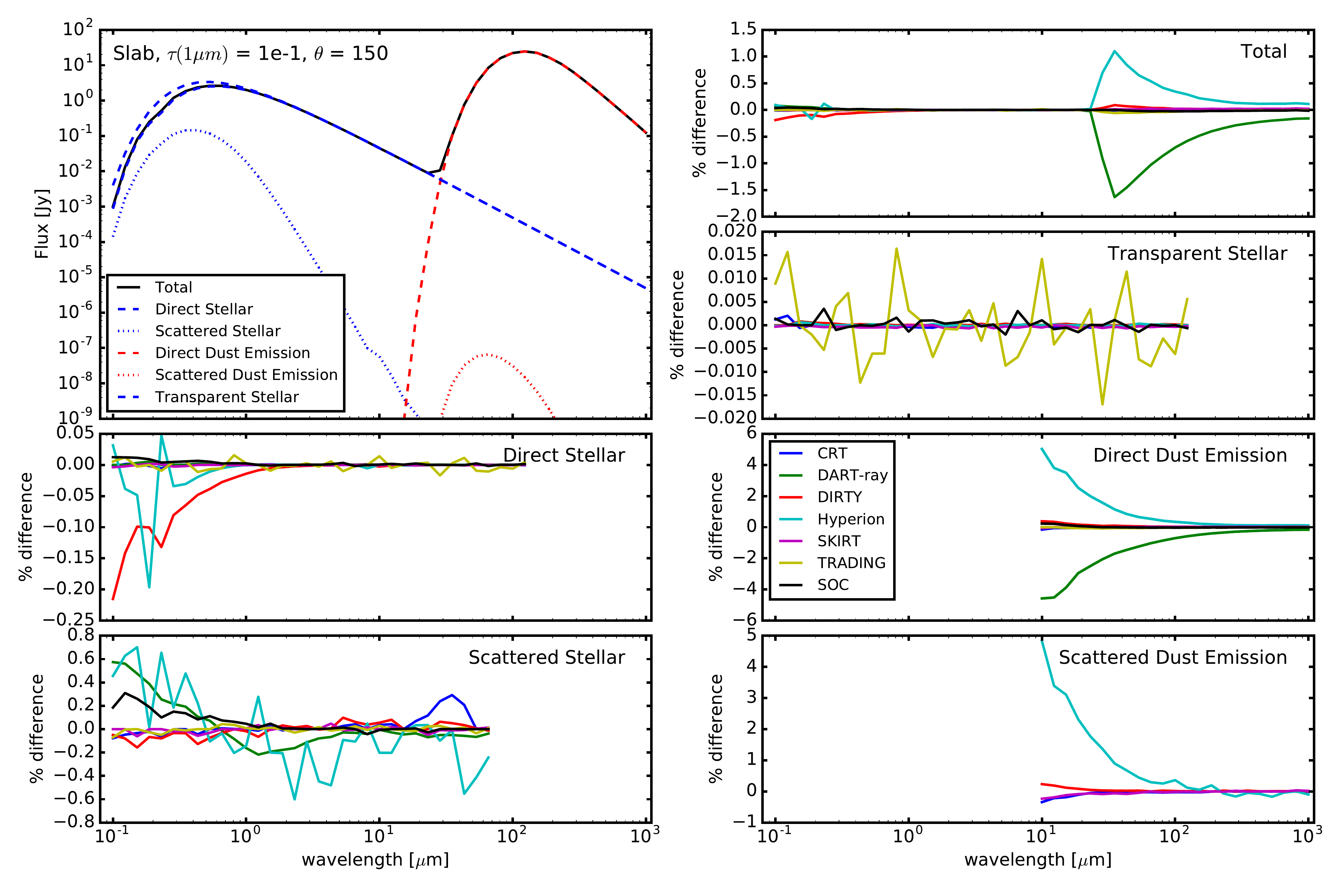}}
\caption{An example of the model global SED outputs are shown for the $\tauref = 0.1$, $\theta = 150^\circ$, and effective grain case.  
The plot in the upper left corner gives the median SED from all the models, both as a total and decomposed into components.  
The other plots give the percentage differences from the median for each of the components.}
\label{fig_example_sed}
\end{figure*}

\begin{figure*}
\resizebox{\hsize}{!}{\includegraphics{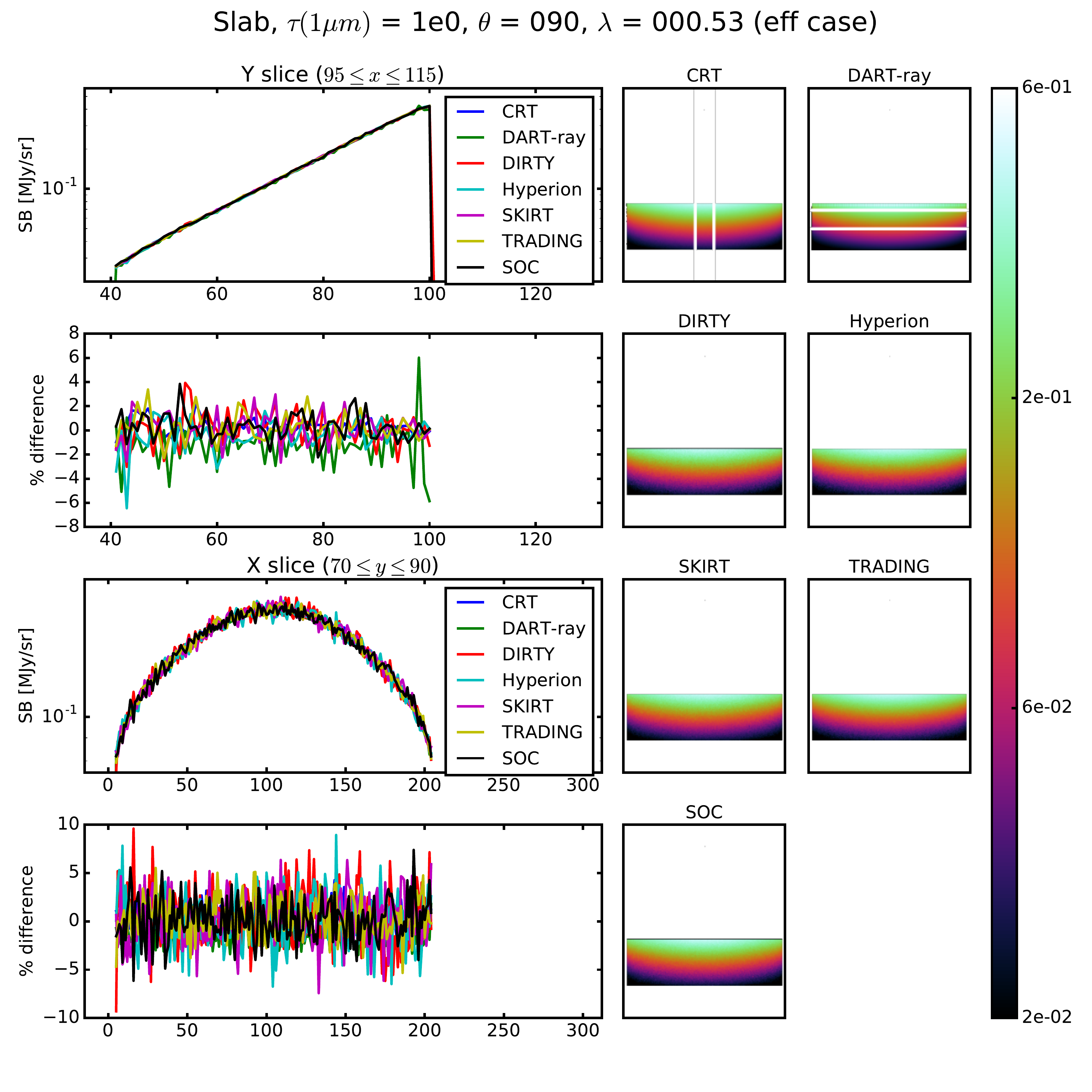}}
\caption{An example of the model image outputs is shown for the $\tauref = 1$ and effective grain case.
In addition to the total images for each model, Y and X slices are shown along with the differences for each model from the median slice.
The X and Y slices refer to the output image dimensions, not the axes in the 3D model space.
The locations of each slice are shown over-plotted on the 1st (Y-slice) and 2nd (X-slice) model images where the slice is computed as an average over the slice width.
}
\label{fig_example_image}
\end{figure*}

The comparisons between the results of the different codes was done both with the global SEDs and images as shown in Figs.~\ref{fig_example_sed} and \ref{fig_example_image}.
The global SED comparisons were done both for the total SED as well as the different components (\S\ref{sec:ex_outputs}) of the RT solution.
The image comparisons included both side-by-side display of the images as well as quantitative comparisons using two slices, one in the X and one in the Y direction.
The slices were averaged in the direction perpendicular to the slice.
The comparison between models was focused on a comparison of the behavior of these averaged slices.
The global SED comparisons were diagnostic of issues with the different components of the solution (e.g., dust scattering and emission).
The image comparisons were diagnostic of general issues in creating the images as well as cases seen only for a limited range of parameters (e.g., UV images at high optical depths).
Quantitative analysis of the images focused on the Y slices as these were diagnostic of systematic issues at all $\theta$ values while being more robust to noise associated with the number of photons/rays.

The comparisons for all cases are given in the Slab section of the TRUST website\footnote{http://ipag.osug.fr/RT13/RTTRUST/BM1.php}.

\subsection{Precision goal}

In contrast with problems that have an analytic solution, for problems such as 3D dust radiative transfer that require a numerical approach, the solution will never converge fully to infinite precision.
Thus, we need to define a robust metric to judge the convergence of the solutions presented here.
One criteria could be when the convergence is well below the expected accuracy of the observations that the models are attempting to reproduce.
Many observations are limited to accuracies of 1\% or larger based on uncertainties in absolute flux or surface brightness calibration \citep[e.g.,][]{Bohlin11, Balog13}.
Another approach is to use the differences seen in previous 2D dust RT benchmark as an upper limit.
The differences between the global SEDs in the \citet{Pascucci04} 2D disk geometry benchmark for all optical depths ($\tau(V) = 0.1 - 100$) was $< 10\%$ and $< 3\%$ for the the lowest optical depth.
The \cite{Pinte09} circumstellar disk geometry benchmark for high optical depths ($\tau(I) \approx 10^3 - 10^6$) found differences between codes on the order of 10\%.
Finally, available computational power imposes a limit - it should be possible to carry out the computations in a reasonable time.
This allows for more codes to participate in this benchmark and makes it reasonable for new codes to use these benchmarks to test their accuracy.
Combining all these points, we adopt a precision goal of 1\%\ on the global SED at lower optical depths and a more relaxed 10\%\ for higher optical depths.
For the Y slice image-based comparisons, we have chosen a 10\%\ precision goal.

\subsection{Resulting precisions}

\begin{table}
\caption{SED average deviations}
\label{tab_sed_dev}
\begin{tabular}{lcc}
\hline\hline
Component & $\tauref \leq 1$ & $\tauref = 10$ \\
\hline
Direct Stellar & 0.3\% & 1.7\% \\
Scattered Stellar & 0.9\% & 3--58\% \\ \hline
\multicolumn{3}{c}{eff} \\ \hline
Direct Dust Emission & 0.7\% & 4.0\% \\
Scattered Dust Emission & 0.7\% & 3.7\%\\ \hline
\multicolumn{3}{c}{equ} \\ \hline
Direct Dust Emission & 0.3\% & 3.1\% \\
Scattered Dust Emission & 1.3\% & 3.3\% \\ \hline
\multicolumn{3}{c}{sto} \\ \hline
Direct Dust Emission & 2.9\% & 3.2\% \\
Scattered Dust Emission & 2.8\% & 2.8\% \\
\hline
\end{tabular}
\end{table}

\begin{table}
\caption{Y Slice average deviations}
\label{tab_yslice_dev}
\begin{tabular}{rcc}
\hline\hline
$\lambda$ (\micron) & $\tauref \leq 1$ & $\tauref = 10$ \\
\hline
0.15 & 5.3\% & $\gg$1000\% \\
0.53 & 3.5\% & 119\% \\ \hline
\multicolumn{3}{c}{eff} \\ \hline
35.11 & 8.3\% & 3.9\%\\
151.99 & 3.4\% & 2.1\% \\ \hline
\multicolumn{3}{c}{equ} \\ \hline
35.11 & 3.8\% & 5.0\%\\
151.99 & 4.7\% & 0.9\% \\ \hline
\multicolumn{3}{c}{sto} \\ \hline
8.11 & 10.7\% & 8.7\%\\
23.10 & 6.7\% & 19.9\% \\
151.99 & 6.5\% & 4.1\% \\
\hline
\end{tabular} \\
\end{table}

The percentage differences between codes for the global SED components are summarized in Table~\ref{tab_sed_dev} and for the Y slices at specific wavelengths in Table~\ref{tab_yslice_dev}.
In general, we find the results to be within the goal precisions with the notable exceptions of the stellar scattered and dust emission components for $\tauref = 10$.
The details of these comparisons are discussed below.

\subsubsection{Stellar radiative transfer}

\begin{figure*}
\resizebox{\hsize}{!}{\includegraphics{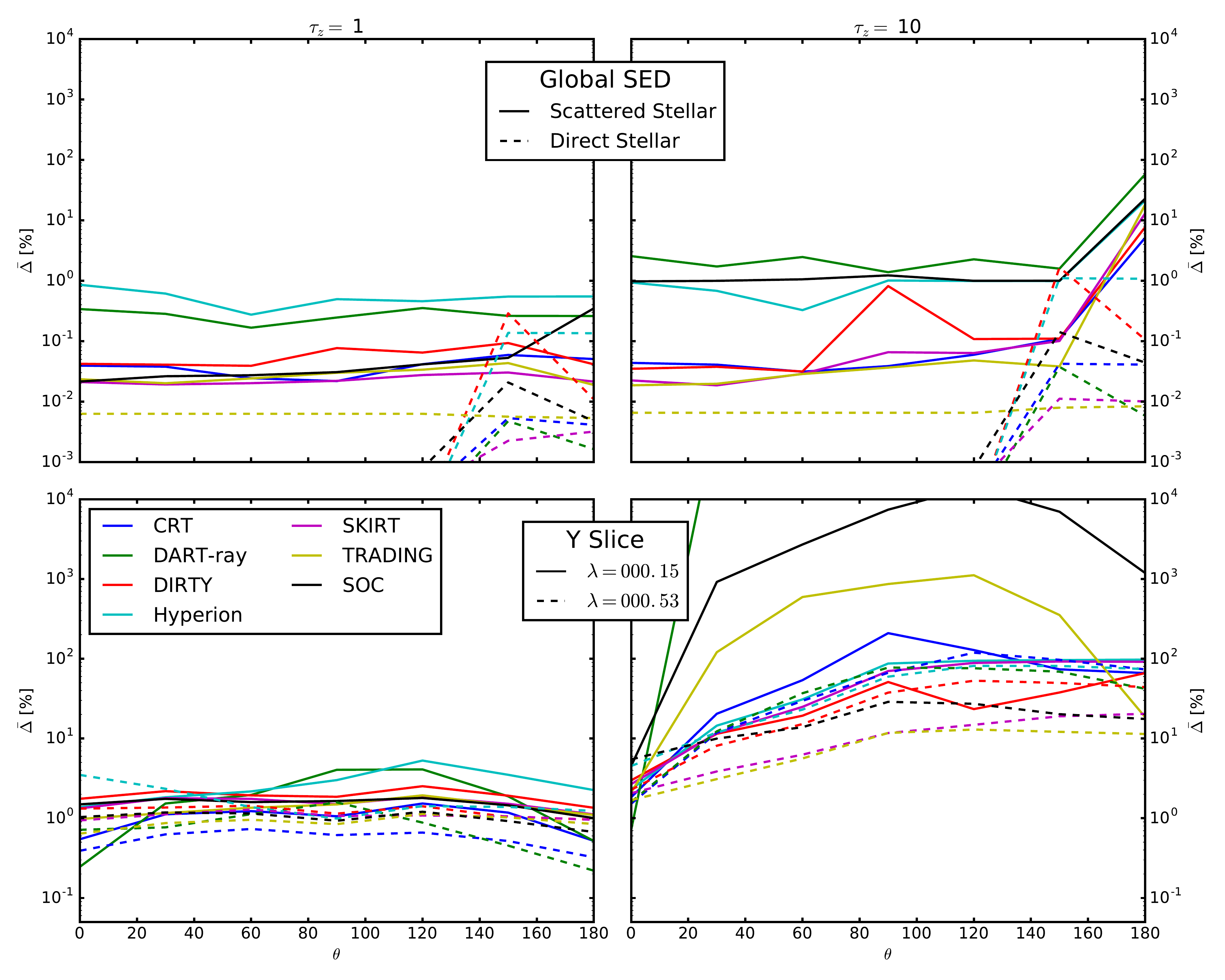}}
\caption{The average deviation of the global SEDs from the median results are shown versus $\theta$ for the direct and scattered stellar photons in the top row.  
The Y slice average deviation from the median results are shown versus $\theta$ for two UV/optical wavelengths probing the scattered stellar photons in the bottom row.
The results for $\tauref = 0.1$ and $0.01$ are similar to those for $\tauref = 1.0$.}
\label{fig_eff_stell_comp}
\end{figure*}

While the properties of the calculated dust emission are sensitive to whether an approximation is used (eff or equ) or
the full solution is calculated (sto) - see Fig.~\ref{fig_example_emission_type} - the radiative transfer problem itself is not.  
In fact, the radiative transfer of photons through dust is mathematically equivalent regardless of whether the photon interaction is computed separately for each size and composition in a distribution, or computed for an effective grain generated by integrating the grain properties over the size distributions and chemical compositions \citep{Steinacker13}. 
Hence, comparisons of the direct and scattered stellar light provide a comparison of each code's treatment of the radiative transfer problem, free from the additional computational complications arising from computing the dust emission. 

The direct and scattered stellar comparisons for the global SEDs are shown in Fig.~\ref{fig_eff_stell_comp} and summarized in Table~\ref{tab_sed_dev}. 
The direct stellar component shows the largest differences at high $\theta$ values where the slab occults the illuminating star.
They are largest at $\theta = 150^\circ$, grow with increasing \tauref, and are just above the goal of 1\% at all optical depths.
The $\theta = 150^\circ$ case provides the maximum optical depth from the star to the observer.
The differences for $\theta < 150^\circ$ are due to different numerical representations of the intrinsic SED between codes and are $\ll 0.1$\%.
For the scattered stellar component, the difference is $<$1\% for $\tauref \leq 1$ and 3\% for $\tauref = 10$ with the exception of $\theta = 180^\circ$ where the difference is 50\%.
Larger differences are not unexpected at $\theta = 180$ as there are no paths for scattered photons that do not require them to penetrate the entire slab, making the results very sensitive to the number of photon packets run for a given model. 
Thus, the $\tauref = 10$ and $\theta = 180^\circ$ is a very sensitive test of the dust scattering at high optical depths.

The comparisons for the Y slices of the images at two representative wavelengths (one in the UV and one in the optical) are shown in the bottom row of Fig.~\ref{fig_eff_stell_comp} and summarized in Table~\ref{tab_yslice_dev}.
For $\tauref \leq 1$, the precisions are well within the goal of 10\%.
At $\tauref = 10$, the discrepancies are much, much larger than this goal.
The plot shows a bifurcated behavior for $\theta > 0^\circ$ that is the signature of very large variations between the median and all of the code results.
The results that are below the median by a large value give $\bar{\Delta} \sim 100\%$ as they are effectively zero.
The results that are above the median by a large value give $\bar{\Delta} \gg 10^4\%$.
For $\theta = 0^\circ$, all the codes have $\bar{\Delta} < 10\%$ as the scattered stellar images are dominated by back scattering off the surface of the slab.
Physically, one would expect that, even for strongly forward-scattering grains at very high optical depth, there would be very little scattered stellar light in the models for viewing angles near $\theta = 180^\circ$.  
Indeed, that is what is observed for this benchmark.  
However, given that the amount of scattered stellar light is very small, small differences in how different codes handle the scattering (and the RT problem in general) can lead to very large relative discrepancies, as observed.  
These results illustrate that 3D dust RT at high optical depth is still a challenging numerical problem for RT codes and can provide a sensitive probe of the efficacy of the numerical solution implemented. 

\subsubsection{Dust emission}

The dust emission changes significantly depending on the assumption used as shown in Fig.~\ref{fig_example_emission_type}.   
The results for each of the dust emission approximations (eff = effective grain, equ = equilibrium only, and sto = full solution including stochastically heated grains) are given.
Almost all of the codes provide results for all three dust emission cases. The exceptions are Hyperion, which only provided results for eff case results, and SOC, which does not provide the equ case results.
In both cases, the codes could compute the missing cases, but time limitations meant they were not computed.

\begin{figure*}
\resizebox{\hsize}{!}{\includegraphics{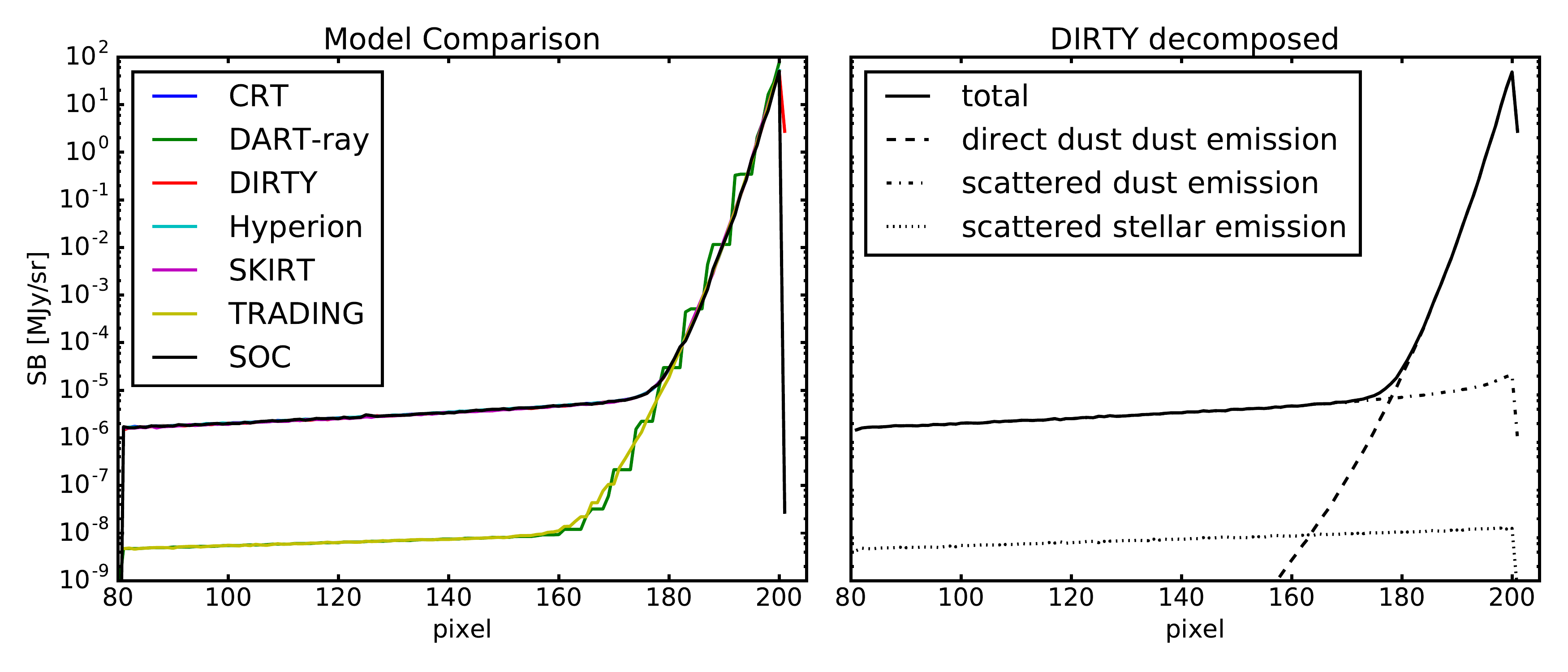}}
\caption{The contribution the scattered dust emission at $\lambda = 35.11~\micron$ makes to the Y slice for $\tauref = 10$, $\theta = 90^\circ$ is shown.
The comparison between all the code results is shown on the left, where the DART-ray and TRADING results are significantly below the results from the other five codes.
On the right, the results from DIRTY are shown decomposed into the three contributing components.
For the back three quarters of the slab, the scattered dust emission dominates over the direct dust emission and scattered stellar emission.
The differences between DART-ray and TRADING and the rest of the codes is due to not calculating the scattered dust emisison component.}
\label{fig_dscat_importance}
\end{figure*}

DART-ray and TRADING do not compute the dust-emission scattered component.
The importance of the dust-scattered emission is illustrated in Fig.~\ref{fig_dscat_importance}.
In addition, DART-ray does not allow for the heating of dust due to its own emission.
The importance of dust self-heating is discussed in \S\ref{sec_self_heating}.
The lack of the full dust emission radiative transfer calculation in these two codes means that it is expected that their results will be less accurate for high optical depths.
High optical depths are where the dust-emission scattering and self-heating are particularly important.
For these reasons and to be conservative, we do not include DART-Ray in determining the precisions of the solution for any optical depth for the global SED dust-emission components.
In addition, we do not include DART-ray or TRADING for the $\tauref = 10$ results for determining the precision of the dust emission Y slices.
We do include all the results in the figures, with those results not used for precision calculations shown as faint lines.
For longer IR wavelengths the results for TRADING do not seem to be significantly affected by not including the dust-emission scattering calculation, but for the purposes of establishing the precision of the benchmark, we conservatively do not include them.

\begin{figure*}
\resizebox{\hsize}{!}{\includegraphics{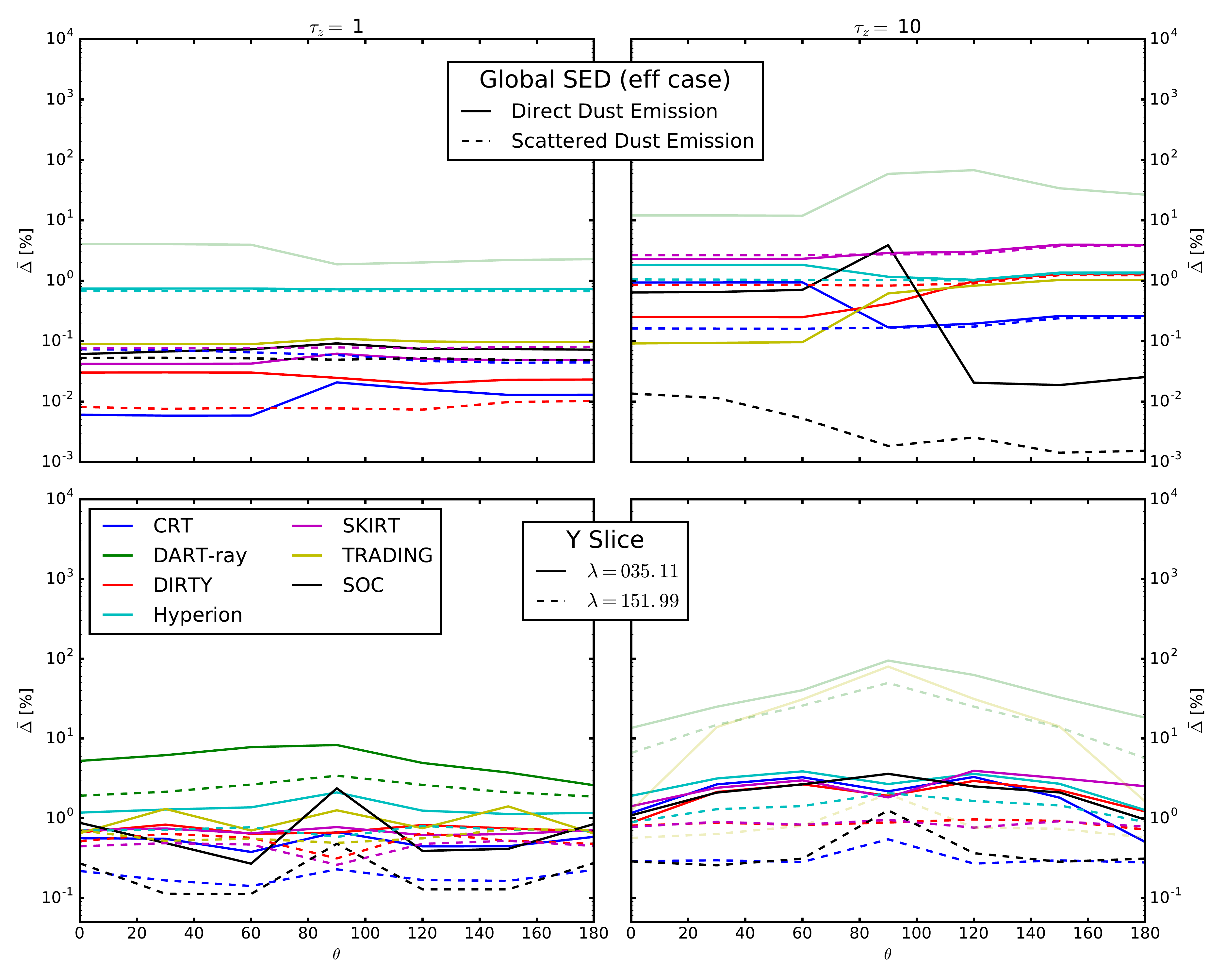}}
\caption{The average deviation from the median results are shown versus $\theta$ for the direct and scattered dust emission for the effective grain approximation in the top row.
The Y slice average deviation from the median results are shown versus $\theta$ for two IR wavelengths probing the dust emission in the bottom row.
The results for $\tauref = 0.1$ and $0.01$ are similar to those for $\tauref = 1.0$.
Models not used in the precision calculation are shown as faint lines.}
\label{fig_eff_demis_comp}
\end{figure*}

\begin{figure*}
\resizebox{\hsize}{!}{\includegraphics{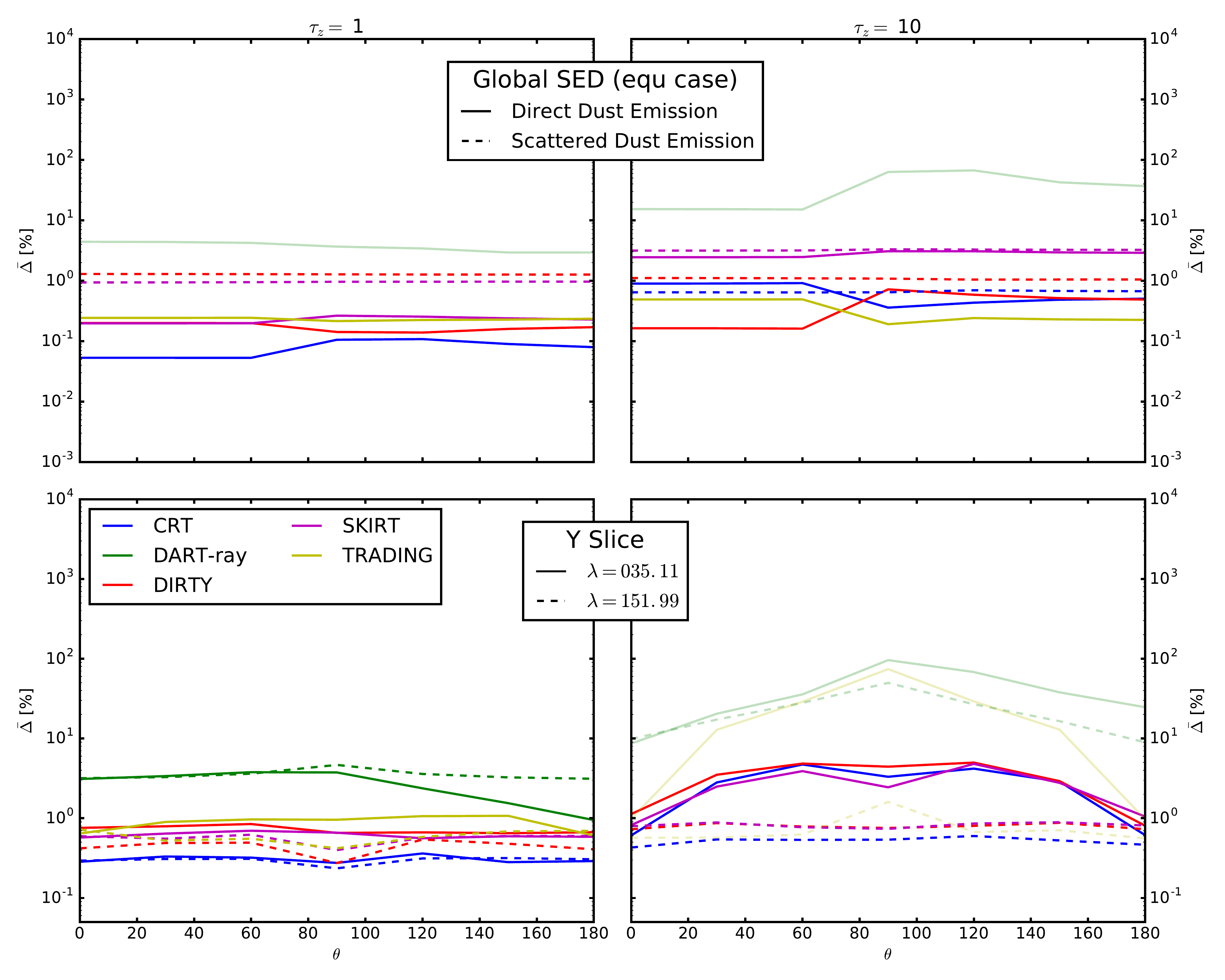}}
\caption{The average deviation from the median results are shown versus $\theta$ for the direct and scattered dust emission for the equilibrium only grain approximation in the top row.
The Y slice average deviation from the median results are shown versus $\theta$ for two IR wavelengths probing the dust emission in the bottom row.
The results for $\tauref = 0.1$ and $0.01$ are similar to those for $\tauref = 1.0$.
Models not used in the precision calculation are shown as faint lines.}
\label{fig_equ_demis_comp}
\end{figure*}

\begin{figure*}
\resizebox{\hsize}{!}{\includegraphics{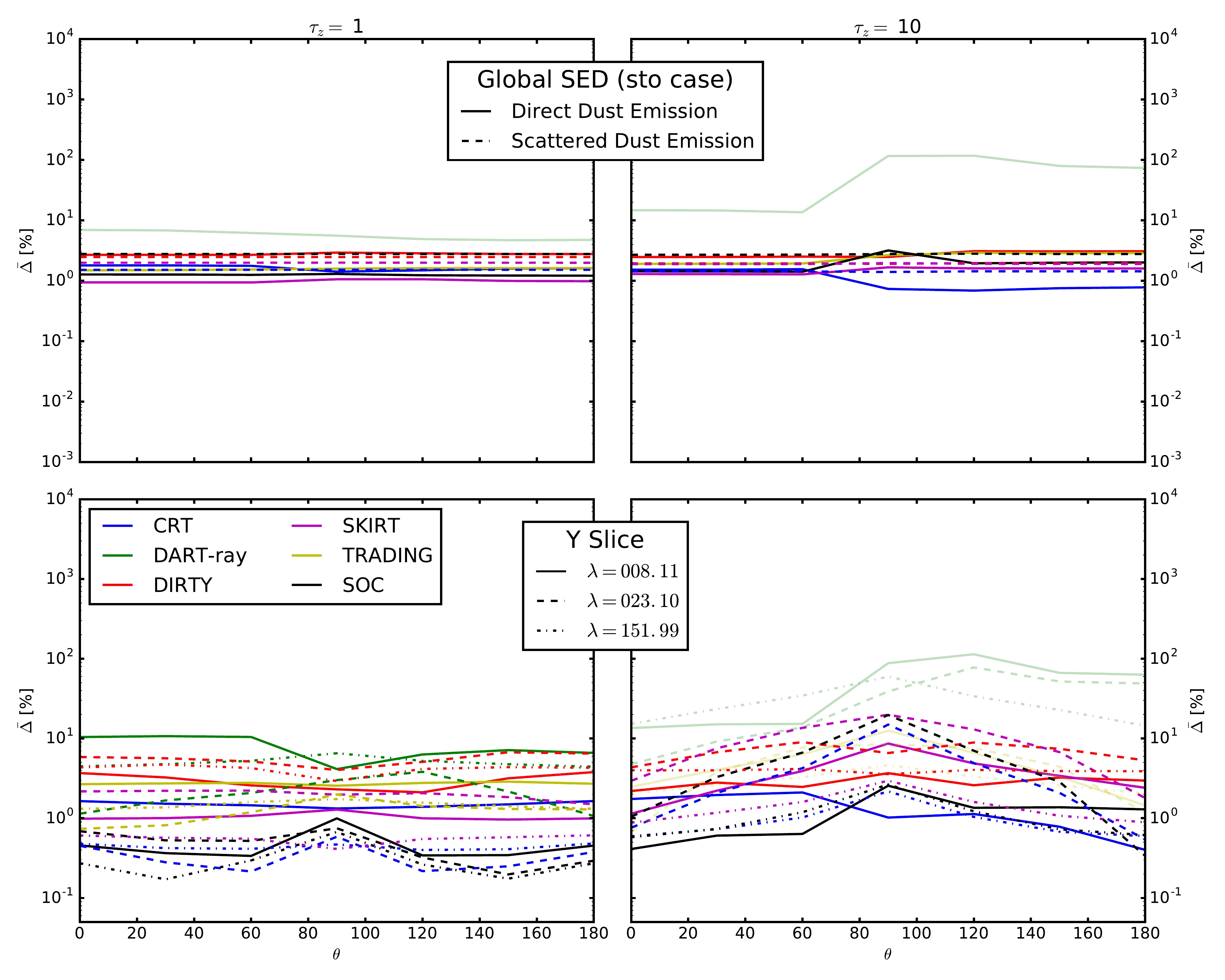}}
\caption{The average deviation from the median results are shown versus $\theta$ for the direct and scattered dust emission for the full grain solution (equilibrium and non-equilibrium dust emission) in the top row.
The Y slice average deviation from the median results are shown versus $\theta$ for three IR wavelengths probing the dust emission in the bottom row.
The results for $\tauref = 0.1$ and $0.01$ are similar to those for $\tauref = 1.0$.
Models not used in the precision calculation are shown as faint lines.}
\label{fig_sto_demis_comp}
\end{figure*}

The direct and scattered dust-emission comparisons for the global SED components are shown in Fig.~\ref{fig_eff_demis_comp}, \ref{fig_equ_demis_comp}, and \ref{fig_sto_demis_comp} for the eff, equ, and sto cases, respectively, and summarized in Table~\ref{tab_sed_dev}.  
The comparisons for the Y slices of the images at two representative wavelengths (one in the mid-IR and one in the far-IR) are shown in the bottom row of the same figures and summarized in Table~\ref{tab_yslice_dev}.
For the eff case, the precisions achieved are at or better than the goals.
For the equ case, the precisions achieved meet the goals, except for the global- dust-emission scattered component that has a precision of 2\%.
For the sto case, the precision was 3\% for the global dust-emission components, well above the goal of 1\%.
The Y slice precisions were better than the goal of 10\%, except for the $\lambda = 23.10$ where the differences were at the 20\% level.

While the goal precisions were achieved for the eff case, it is worth noting that the differences between the Hyperion results and most of the codes are likely caused by differences in the sampling of the photons' wavelengths.
Specifically, most of the codes sample the photon wavelengths directly from the wavelength grid while Hyperion samples the photon wavelengths from a continuous spectrum and the wavelength grid is only used for output quantities.
With a higher-resolution wavelength grid, it is likely that most of the codes would cluster around the Hyperion results.

\section{Convergence tests}
\label{sec_convergence}

\begin{table}
\caption{DIRTY base parameter values}
\label{tab_model_params_converge}
\begin{tabular}{lcl}
\hline\hline
Name & Values & Description \\
\hline
$N$ & $10^8$ & \# of photons per wavelength \\
$n_{xy}$ & 100 & \# of bins in $x$ or $y$ \\
$n_z$ & 100 & \# of bins in $z$ \\
$m_\mathrm{scat}$ & 500 & max \# of scatterings \\
$n_\mathrm{iter}$ & 4 & \# of dust heating iterations \\
$\xi_\mathrm{scat}$ & 0.5 & scattering composite bias \\
$\xi_\mathrm{emis}$ & 0.5 & emission composite bias \\
\hline
\end{tabular}
\end{table}

In the absence of an analytic solution, another method for building confidence in the numerical solution is to perform convergence tests. 
Such tests also provide insight into the effects different limits place on the solution (e.g., the importance of scattered photons or dust self-heating).
These tests involve changing numerical tolerances and quantifying how the solution changes.
Convergence testing is most often done based on an experience-based understanding of the relative importance of the model parameters in influencing the solution.

We have performed a number of quantitative convergence tests for this benchmark using the DIRTY code. 
Similar results would be expected for the other Monte Carlo codes and, to a lesser extent, with Ray-Tracing codes.
For the convergence tests, the parameters not being tested were set to values given in Table~\ref{tab_model_params_converge}.  
We have not exhaustively searched the possible parameter space in our convergence tests, due to the significant computational resources necessary for each set of parameter tests.  
Instead, we have fixed all the parameters, except the one being varied, to values that are expected to provide reasonable precision based on previous runs.

We have performed the convergence tests for $\tauref = 1$ and 10.
Practically, we found that testing the precision for $\tauref = 1$ could be done in a reasonable amount of computer time and the results for lower optical depths will have comparable or better precision.
The $\tauref = 10$ was challenging for all codes and the precision remains limited by computer time (i.e., $< 1$ month or so single threaded being reasonable within the scope of this study).

\subsection{Number of photons/rays}

The number of photons or rays that are computed at each wavelength ($N$) is clearly a parameter that will strongly influence the precision of the model results.
This model parameter controls the precision of the scattered light calculation and of the dust emission from each grid cell.
A significant portion of the computations scale directly with $N$. 

\begin{figure*}
\resizebox{\hsize}{!}{\includegraphics{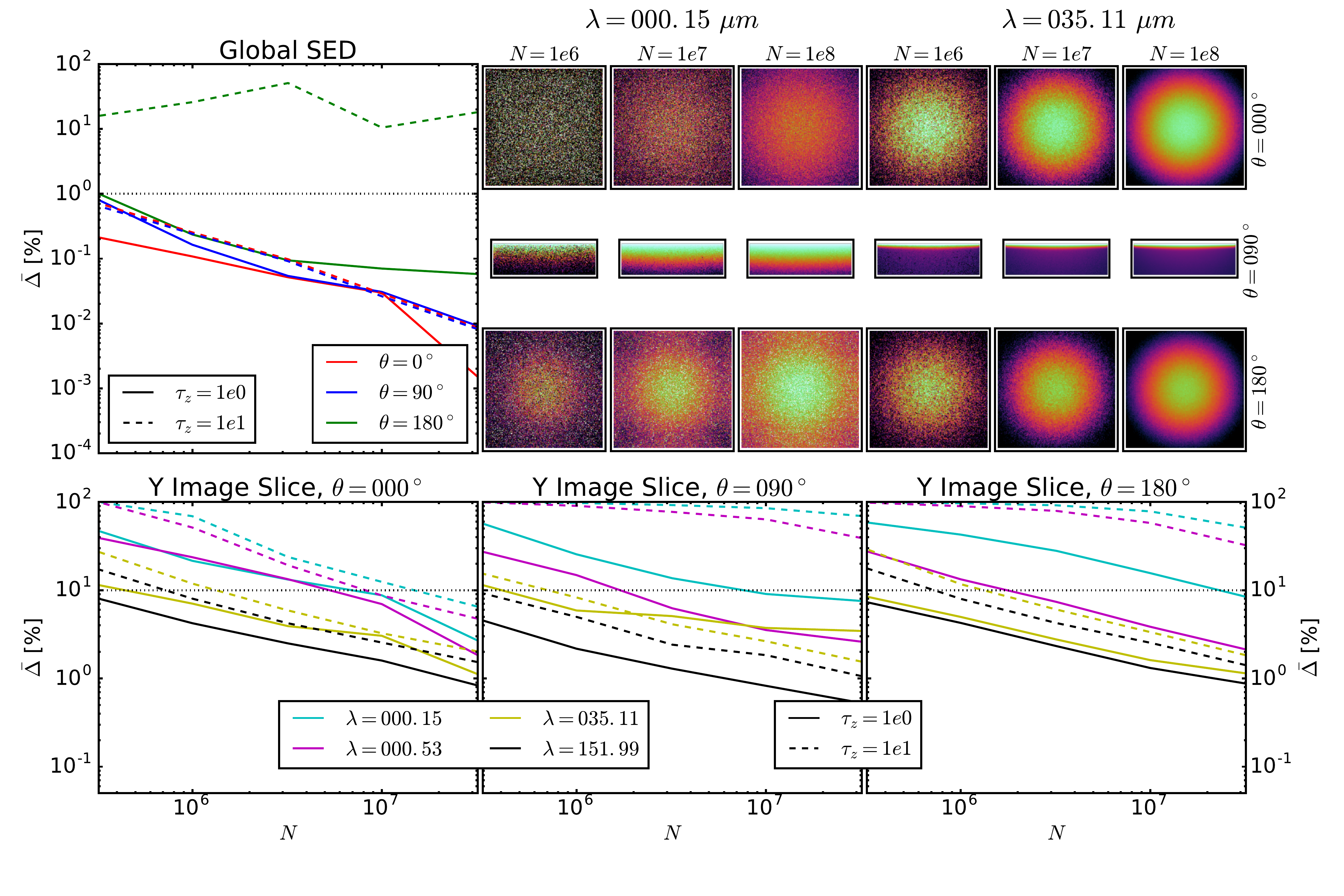}}
\caption{The average deviations ($\bar{\Delta}$) versus the number of photons or rays that are computed at each wavelength ($N$) are shown for the total global SEDs and Y image slices.  
The images are for the most challenging case of $\tauref = 10$ and illustrate the qualitative impact of increasing $N$.
Plots and images are shown for $\theta = 0^\circ$, $90^\circ$, and $180^\circ$.  
The dotted horizontal lines give the 1\% (global SED) and 10\% (Y slices) levels.
The images are plotted with the same log scaling for each wavelength and angle combination.
The results for $N = 10^8$ are not shown as $\bar{\Delta}$ is computed relative to this case where $\bar{\Delta} = 0$ by definition.}
\label{fig_converge_nphot}
\end{figure*}

The $N$ convergence tests were computed for three angles to illustrate the impact of $N$ on backscattered photons ($0^\circ$), penetration depth into the slab ($90^\circ$), and penetration through the entire slab ($180^\circ$). 
Fig.~\ref{fig_converge_nphot} plots the average deviation as a function of $N$ for the total global SED and the Y slices.  
The average deviation is computed compared to the model run with the largest $N$ (e.g., $10^8$). 
For $\tauref = 1$, convergence of the global SED and component SEDs (not shown) to 1\% is achieved with $N = 10^6$  and $N = 10^7$, respectively.
For $\tauref = 10$, a similar behavior is seen except for the scattered stellar component where convergence is not seen and the average deviation remains high ($\sim$20-80\%) for all values of $N$ tested. 
The convergence for the Y slices is more complicated.
Convergence to 10\%  is reached by $N = 2 \times 10^7$ for small angles (e.g., $0^\circ$) at all wavelengths.  
For $\tauref = 1$, the goal convergence is seen for the IR wavelengths around $10^6$ and for the UV/optical wavelengths at $10^7$.
For $\tauref = 10$, the goal convergence is again seen for the IR wavelengths around $10^6$, but the convergence for the UV/optical wavelengths is well beyond the number of photons tested with an extrapolated prediction of convergence around $N > 10^9$ photons.

From these calculations, we can see that $N = 10^8$ will provide good precision for all but the UV/optical wavelengths for the $\tauref = 10$ case.

\subsection{Spatial grid}

Another obvious model configuration parameter to test for convergence is the number of spatial bins that describe the slab.
The computation time required for a model scales with the number of bins through the need to compute the dust emission from each bin as well as computing the radiative transfer through all the bins.  
The finer the slab is divided, the more exact the solution becomes, but also the more time required for the computations.
It is useful to note that using the Monte Carlo solution technique with a uniform dust density results in the scattering component of the solution being independent of the number of spatial bins used.
This is because the scattering location is computed exactly for each photon packet independent of the grid.

The slab geometry naturally lends itself to simple division into bins in the $x$, $y$, and $z$ dimensions. 
The convergence tests were done using linear $x$ and $y$ spacing and logarithmic $z$ spacing starting from the slab front facing nearest the star.  
We found that logarithmic bin spacing in $z$ provided equivalent results to linear $z$ bin spacing but required fewer bins for the same precision.  

\begin{figure*}
\resizebox{\hsize}{!}{\includegraphics{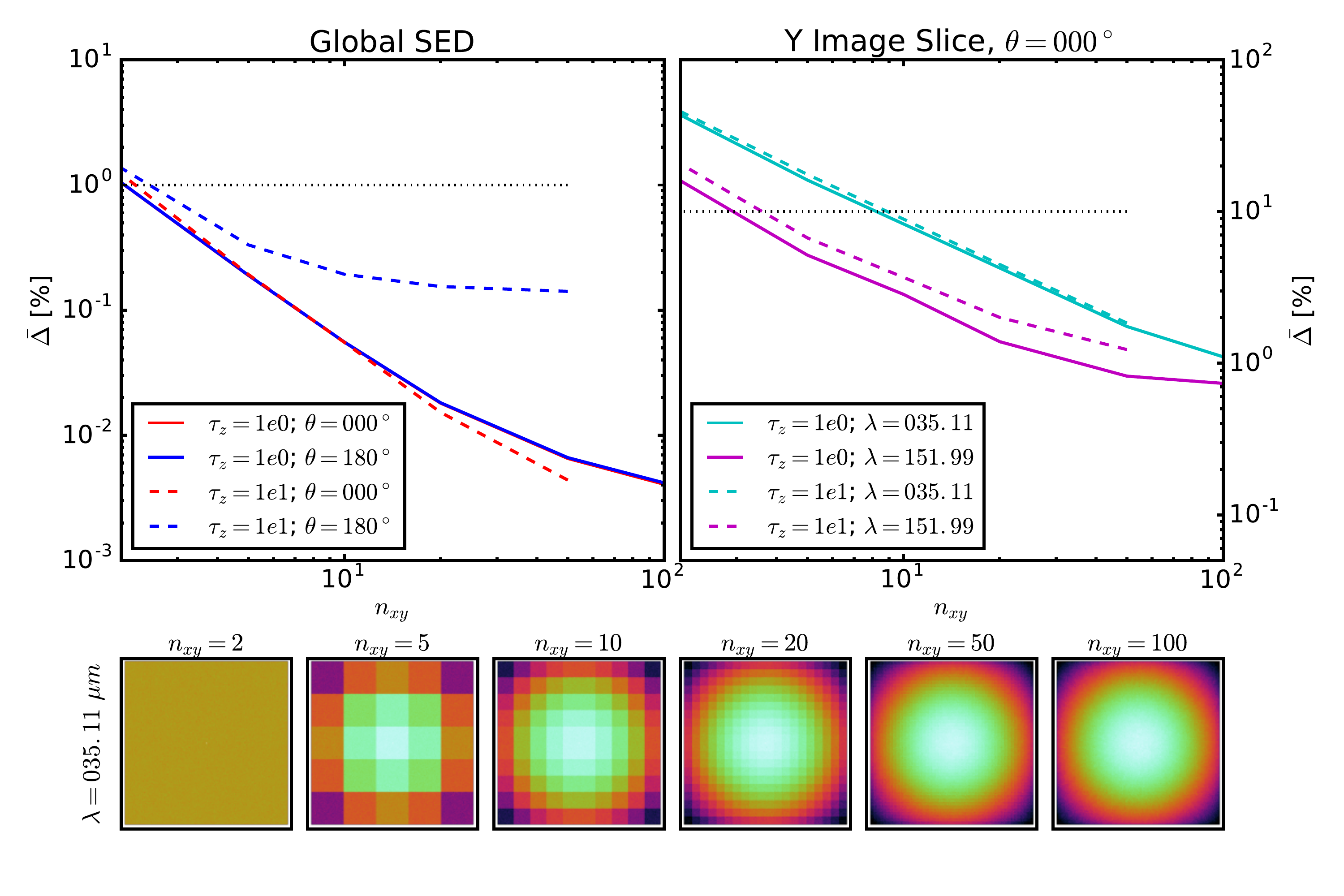}}
\caption{The average deviations ($\bar{\Delta}$) versus $n_{xy}$ are shown for the total global SEDs and Y image slices for $\tauref = 1$ and $10$, and $\theta = 0^\circ$ and $180^\circ$.
The images are for the $\tauref = 1$ and $\theta = 0^\circ$ case and illustrate the qualitative impact of increasing $n_{xy}$.
The images are log scaled over the same range.
The image Y slice results are shown only for $\theta = 0^\circ$ as the results for $\theta = 180^\circ$ are very similar.
Only the image slices at two diagnostic infrared wavelengths probing the dust emission are shown as the ultraviolet and optical scattered light images are not sensitive to $n_{xy}$ for Monte Carlo codes.
The dashed and dotted horizontal lines give the 1\% (global SED) and 10\% (Y slice) lines.
The results for $n_{xy} = 200$ ($\tauref \leq 1$) and $n_{xy} = 100$ ($\tauref = 10$) are not shown as $\bar{\Delta}$ is computed relative to these cases where $\bar{\Delta} = 0$ by definition.}
\label{fig_converge_nxy}
\end{figure*}

The convergence tests for the number of $x$ and $y$ bins ($n_x = n_y \equiv n_{xy}$) were computed for the $\theta = 0^\circ$ and $180^\circ$ cases as these two viewing angles are the most sensitive to $n_{xy}$.  
Fig.~\ref{fig_converge_nxy} plots the average deviation as a function of $n_{xy}$ for the total global SEDs and Y slices. 
The average deviation is computed compared to the model run with the largest $n_{xy}$. 
Convergence of the global SEDs to 1\% is achieved with $n_{xy} = 2$ for the total and all the components.
Similarly, 10\% convergence for the image slices is achieved by $n_{xy} = 10$.

\begin{figure*}
\resizebox{\hsize}{!}{\includegraphics{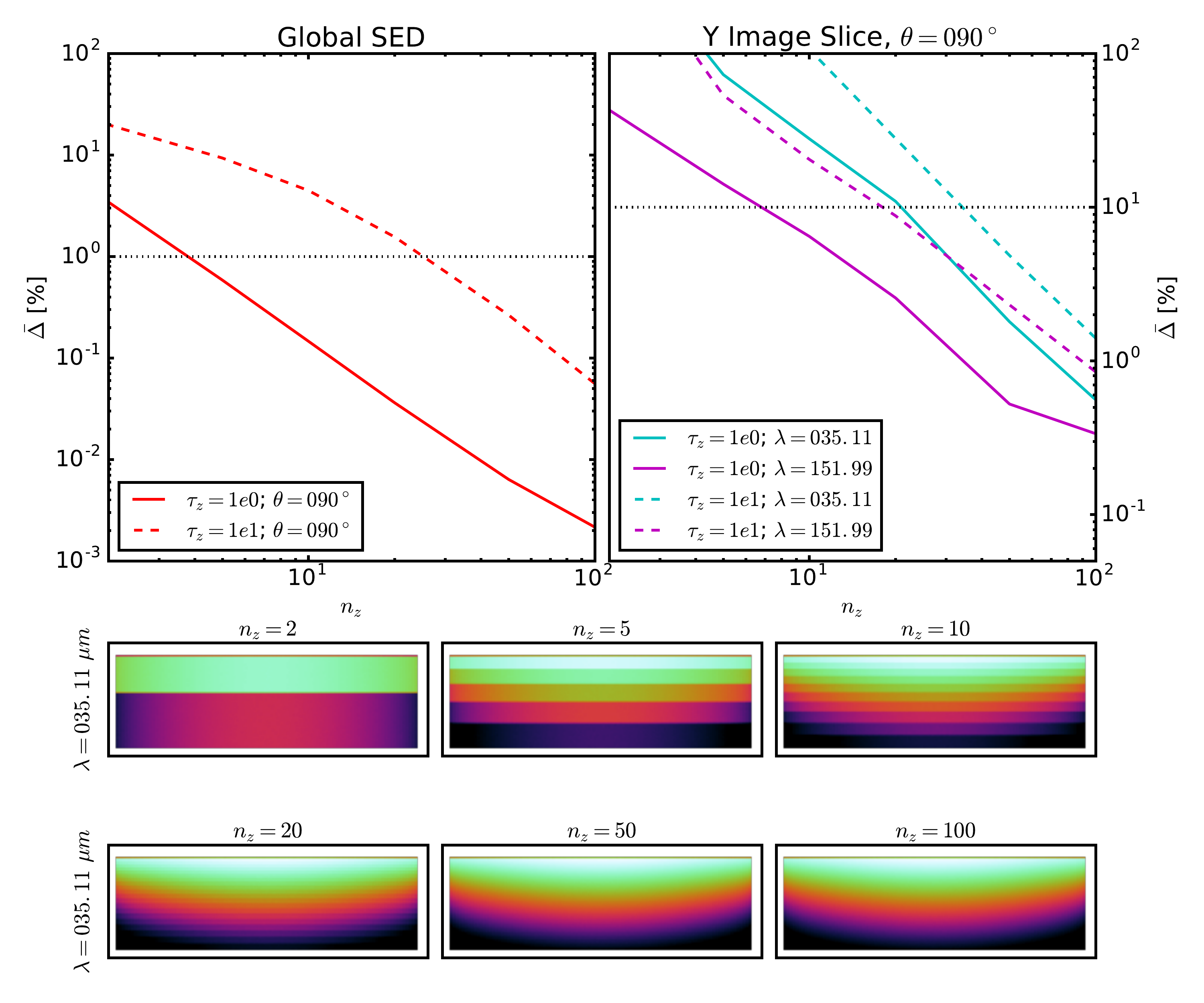}}
\caption{The average deviations ($\bar{\Delta}$) versus $n_{z}$ are shown for the total global SEDs and Y image slices for  $\tauref = 1$ and $10$ and $\theta = 90^\circ$.
The images are for the $\tauref = 1$ and $\theta = 90^\circ$ case and illustrate the qualitative impact of increasing $n_{z}$.
The images are log scaled over the same range.
The DIRTY spatial grid uses a log spacing along the z axis and this is reflected in the images shown.
Only the image slices at two diagnostic infrared wavelengths probing the dust emission are shown as the ultraviolet and optical scattered light images are not sensitive to $n_{z}$ for Monte Carlo codes.
The dashed and dotted horizontal lines give the 1\% (global SED) and 10\% (Y slice) lines.
The results for $n_z = 200$ are not shown as $\bar{\Delta}$ is computed relative to this case where $\bar{\Delta} = 0$ by definition.}
\label{fig_converge_nz}
\end{figure*}

The $n_z$ convergence tests were computed for $\theta = 90^\circ$ as this viewing angle is the most sensitive to $n_z$. 
Fig.~\ref{fig_converge_nz} plots the average deviation as a function of $n_{z}$ for the total global SEDs and Y slices where average deviation is computed compared to the model run with the largest $n_{z}$.  
For $\tauref = 1$, convergence of the global SEDs to 1\% is achieved with $n_{z} \sim 4$.  
Convergence to 1\% for the components of the global SEDs is achieved for $\tauref = 10$ at $n_z \sim 30$.
The image slice convergence to 10\% is achieved at $n_z \sim 20$ for $\tauref = 1$ and $n_z \sim 35$ for $\tauref = 10$.

\subsection{Number of scatterings}

\begin{figure*}
\resizebox{\hsize}{!}{\includegraphics{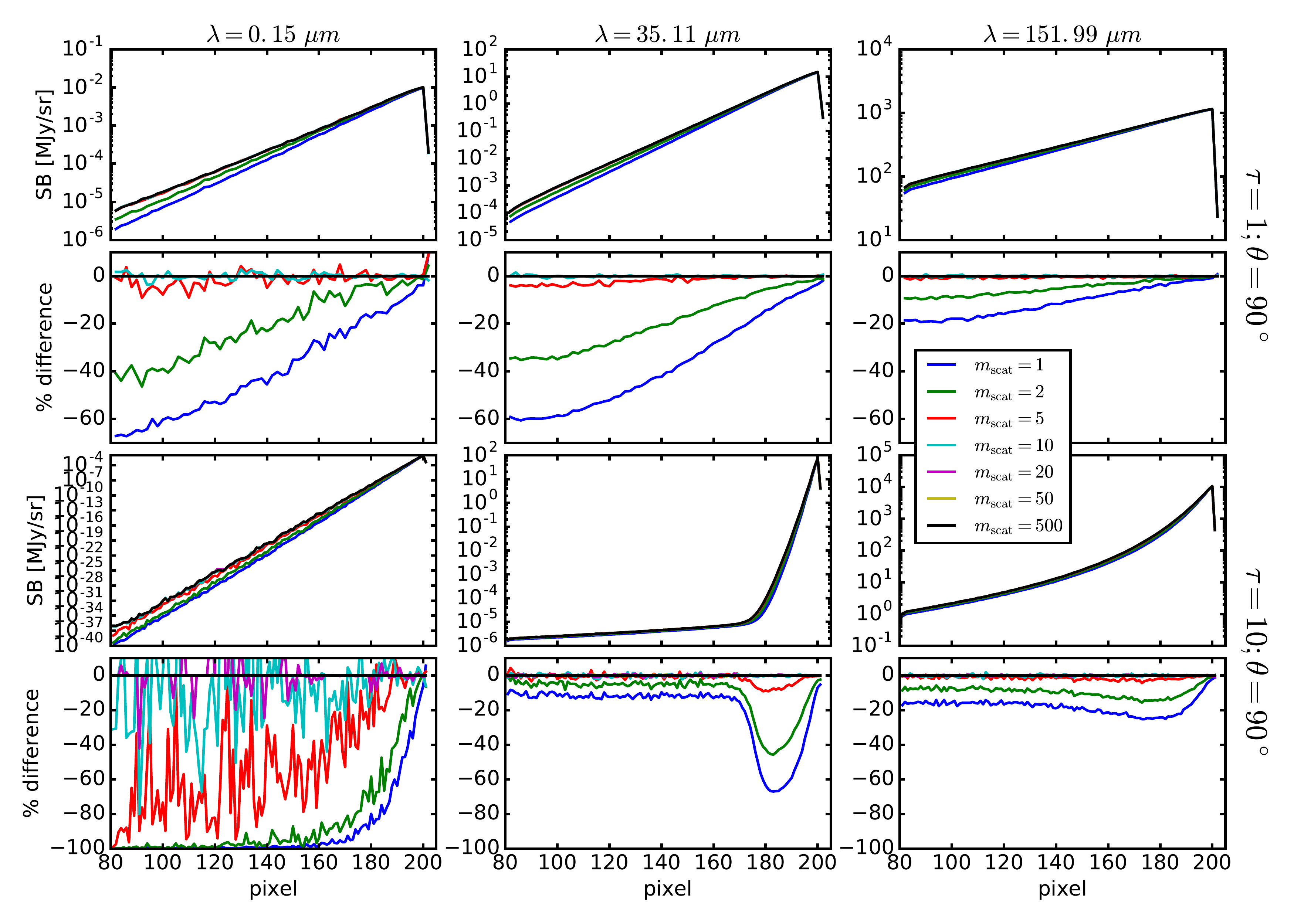}}
\caption{The $\lambda = 0.15$ (left), $35.11$ (center) and $151.99~\micron$ (right) Y slices for $\theta = 90^\circ$ and $\tauref = 1$ and 10 are plotted for a range of maximum allowed number of scatterings ($m_\mathrm{scat}$).}
\label{fig_converge_mscat}
\end{figure*}

Calculation of the scattered photons is one of the challenging parts of the radiative transfer solution, especially in light of the importance of multiple scattered photons at higher optical depths.  
In general, RT codes set a "maximum number of scatterings" to avoid photon packets getting "stuck", especially in very high optical depth environments. 
Of course, if that maximum value is set below the typical number of scatterings a photon might undergo, it can lead to erroneous results from the simulation for the scattered signal.
To quantify the importance of multiple scattering, we carried out convergence tests at $\theta$ values of $0^\circ$, $90^\circ$, and $180^\circ$.
For $\tauref \leq 1$, the number of scatterings needed to achieve the goal precisions is on the order of 5.
For $\tauref = 10$, approximately 20 scatterings are needed for the goal precisions with this driven mainly by the convergence at the shortest wavelengths.
We illustrate the importance of multiple scattering for the $\tauref = 1$ and 10 cases in Fig.~\ref{fig_converge_mscat}.
This figure gives the Y slices and percentage deviations from the $m_\mathrm{scat} = 500$ case for $\lambda = 0.15$, $35.11$, and $151.99$~\micron.
We only show the $90^\circ$ case as it clearly illustrates the changing importance of multiple scattering as a function of penetration depth in the slab.
Multiple scattering is important for both the direct scattering at $\lambda = 0.15$~\micron\ and the dust emission at both IR wavelengths.
The dependence of the IR wavelengths on the number of scatterings clearly shows the importance of dust heating due to scattered photons.
As expected, the dependence on multiple scattering is largest at higher optical depths.

\subsection{Self-heating iterations}
\label{sec_self_heating}

\begin{figure*}
\resizebox{\hsize}{!}{\includegraphics{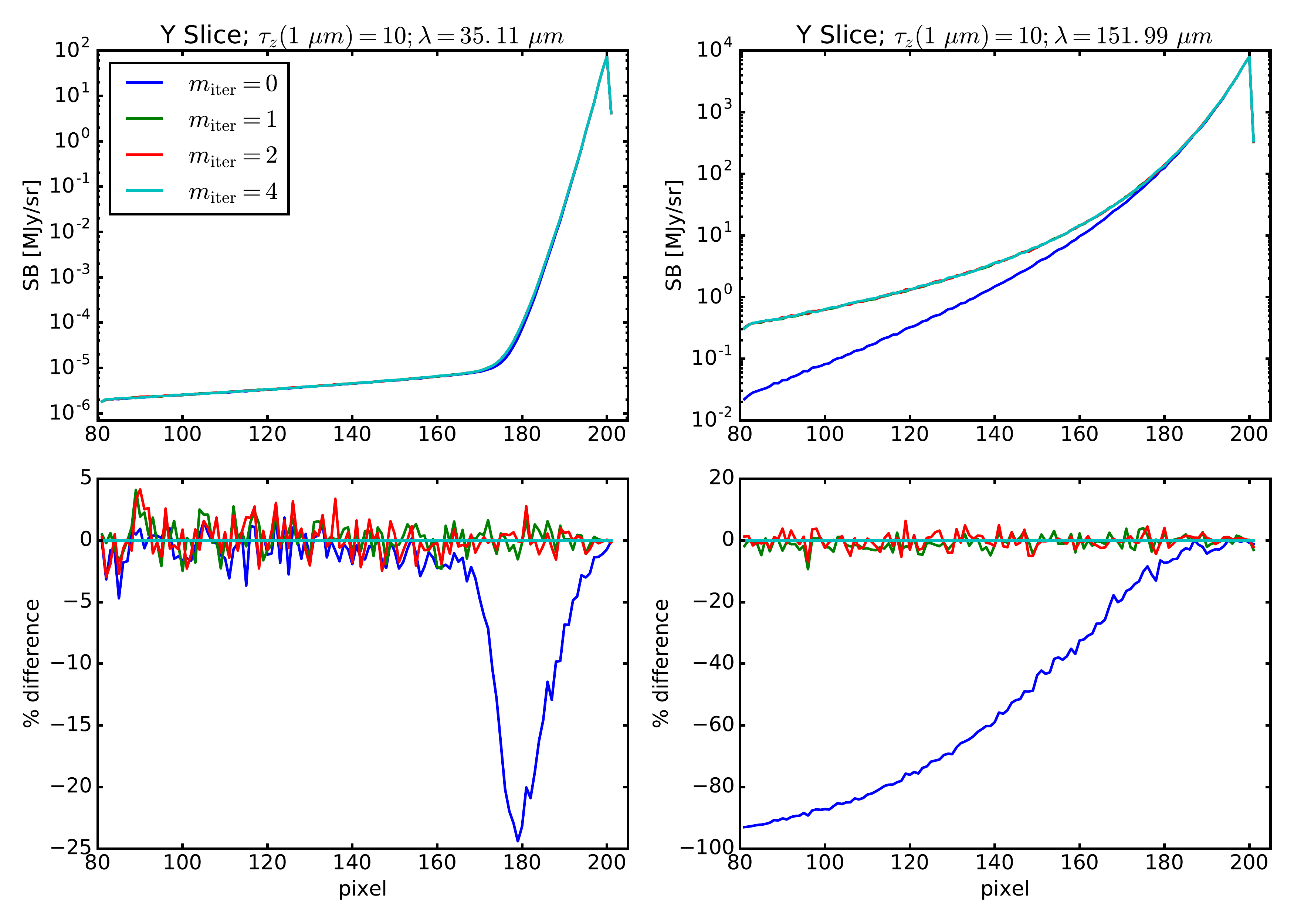}}
\caption{The $\lambda = 35.11$ (left) and $151.99~\micron$ (right) Y slices for $\tauref = 10$ and $\theta = 90^\circ$ are plotted for a range of dust self-heating iterations ($m_\mathrm{iter}$).}
\label{fig_converge_miter}
\end{figure*}

The thermalized radiation emitted by dust that is heated by the primary radiation source can in turn be absorbed by other dust grains in the model space, leading to dust self-heating.
This self-heating increases in importance as the optical depth increases, since, depending on the location, the dominant radiation source may be the re-emitted dust emission. 
Most RT codes account for dust self-heating by iterating between the dust emission and the dust absorption and scattering, stopping when a preset energy convergence is achieved.
We carried out convergence tests to quantify the importance of dust self-heating.
For $\tauref \leq 1$, both the global SED and Y-slice comparisons show that no iterations are necessary to meet our precision goals; the effect of dust self-heating is small enough that neglecting it does not change the precision of the resultant model. 
This is not the case for $\tauref = 10$ where neglecting dust self-heating results in global SEDs in error larger than the goal precisions for all cases and the Y slices for the $\theta = 90^\circ$ case.
The $\theta = 90^\circ$ case clearly shows the impact of dust self-heating and we illustrate this in Fig.~\ref{fig_converge_miter}.
This figure gives the Y slices for a range of dust self-heating iterations starting with no self-heating ($m_\mathrm{iter} = 0$).
The impact of dust self-heating is to raise the emission in the front of the slab at shorter wavelengths (e.g., $\lambda = 35.11~\micron$) and, more dramatically, over most of the slab at longer wavelengths (e.g., $\lambda = 151.99$).
Fortunately, only a single dust self-heating iteration is needed, with additional self-heating iterations providing only small gains.

\subsection{Scattering at high optical depths}
\label{sec_high_tau_scat}

\begin{figure*}
\resizebox{\hsize}{!}{\includegraphics{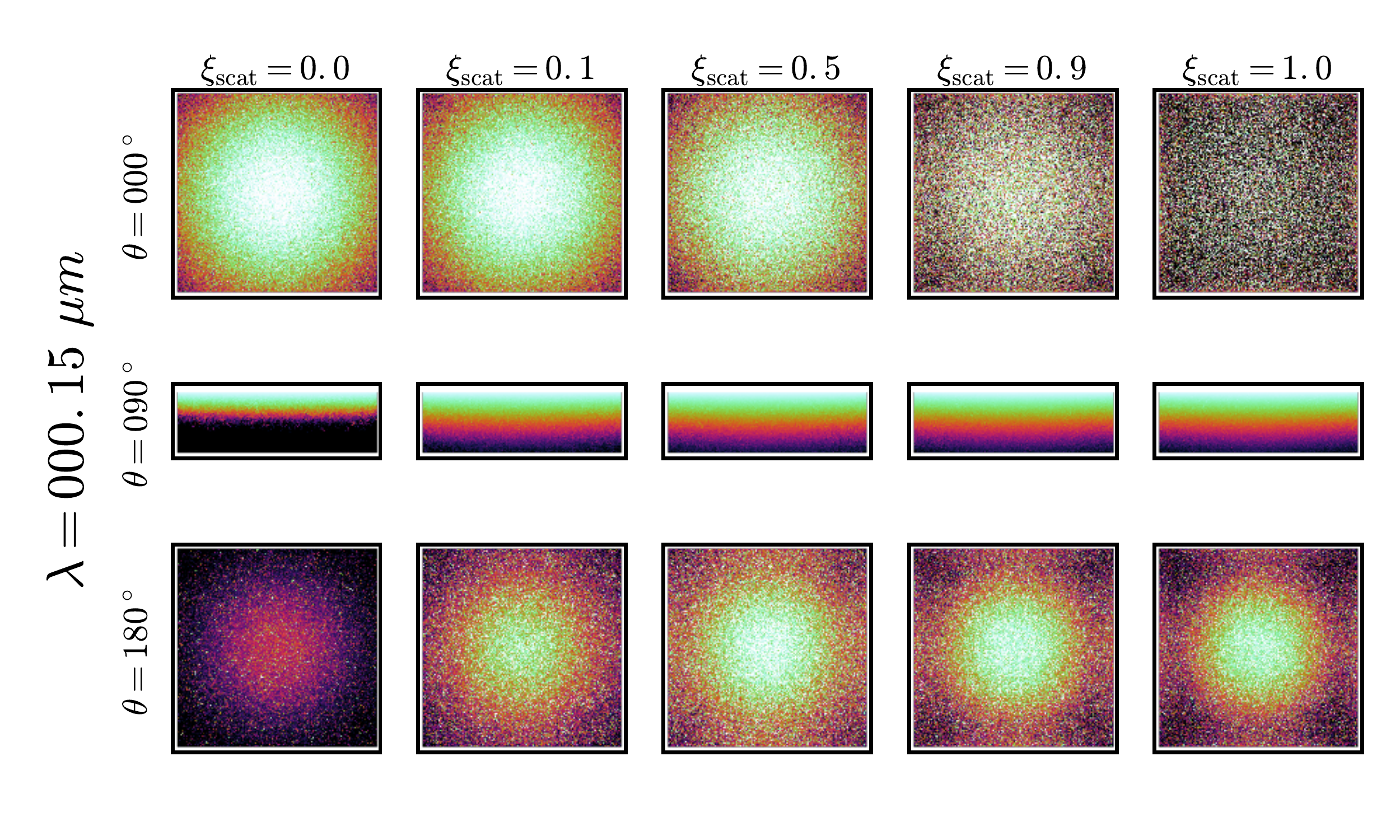}}
\resizebox{\hsize}{!}{\includegraphics{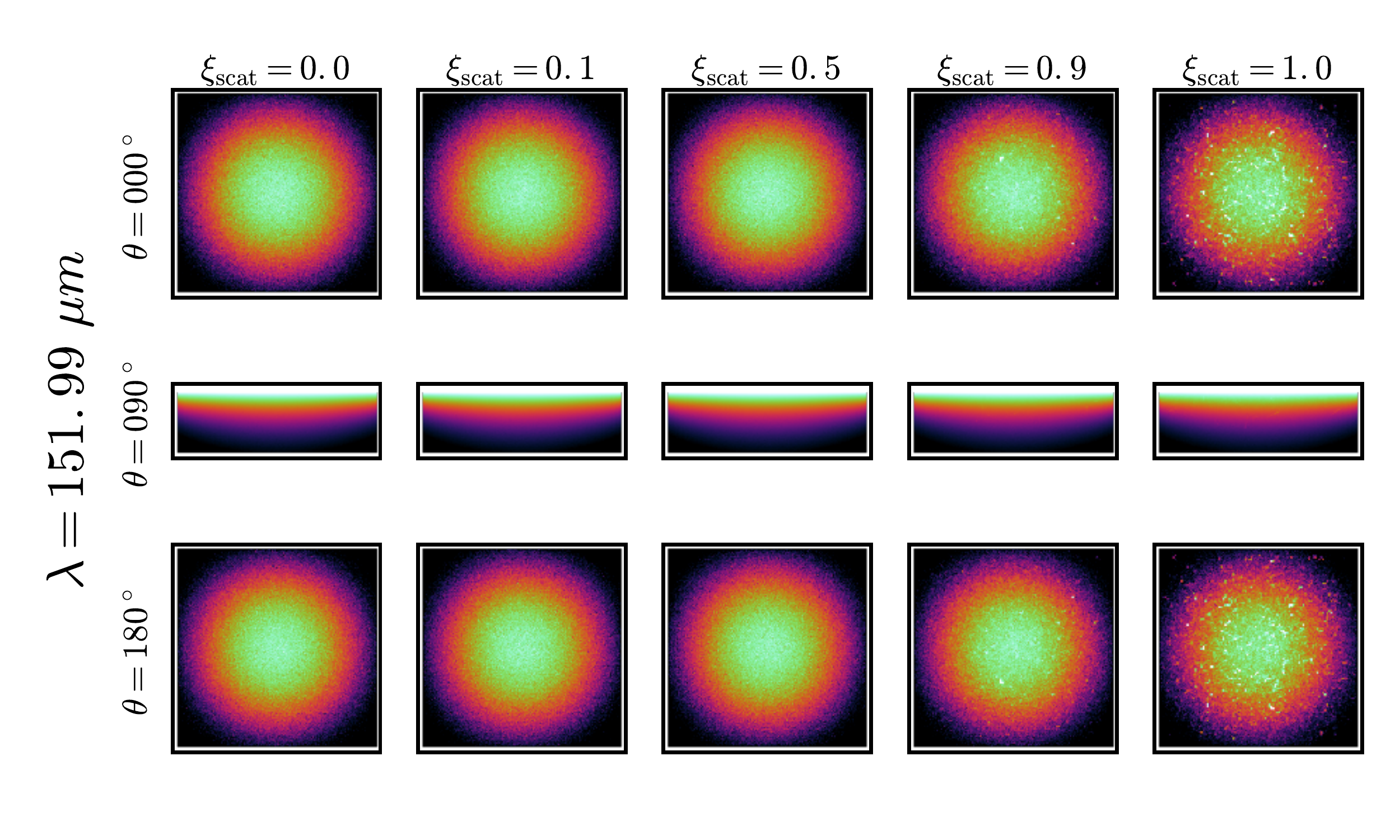}}
\caption{Images in the UV and IR are shown for $\tauref = 10$, $\theta = 0^\circ$, $90^\circ$, and $180^\circ$, and a range of $\xi_\mathrm{scat}$ values.
The images are log scaled and share the same scaling at each unique combination of $\lambda$ and $\theta$.}
\label{fig_xiscat}
\end{figure*}

Radiative transfer at high optical depths through dust is challenging for most numerical solution techniques.
There are approximations possible such as the diffusion approximation \citep{Kuiper2010}, but such approximations impose real limitations on the accuracy of the resulting calculations. 
Motivated by the work on this benchmark, a new technique based on composite biasing was introduced to the Monte Carlo 3D dust RT community by \citet{Baes16}.
The composite bias technique provides a way to sample two probability distributions while controlling the amplification of the resulting photon weight.
Basically, the site of the next scattering is chosen from one of two different distributions with the frequency with which each distribution is used is controlled by the parameter $\xi_\mathrm{scat}$ that varies between 0 and 1.
The first distribution is the standard $e^{-\tau}$ and the second is a much flatter distribution (e.g., a uniform distribution $\tau$).
For example, if $\xi_\mathrm{scat} = 0.5$, then one half the time the first distribution is used and the other half the second is used.
The weight of the photon is modified to account for the difference of the composite of the two distributions from the standard $e^{-\tau}$ distribution.
The $\xi_\mathrm{scat} = 0$ case corresponds to the standard $e^{-\tau}$ distribution.

The need for such a composite technique can be illustrated by considering the $\tauref = 10$ and $\theta = 90^\circ$ case.
For the standard scattering distribution ($\xi_\mathrm{scat} = 0.0$), the back of the slab was approximately $10^{-100}$ fainter than the front of the slab for the $0.15~\micron$ image.
The approximate difference between the scattered light component from the front to back of the slab can be calculated by the simple approximation that intensity from singly scattered photons should be the albedo multiplied by $e^{-\tau}$.
Assuming the scattered light at the front of the slab has $\tau = 0$ and the back of the slab has $\tau \sim 79$ (at $0.15~\micron$ for $\tauref = 10$), then the ratio should be $e^{-79} = 1.8 \times 10^{-35}$.
This is much, much higher than seen using the standard scattering prescription.

Fig.~\ref{fig_xiscat} illustrates the results for a range of $\xi_\mathrm{scat}$ at two representative wavelengths for the $\tauref = 10$ model.  
For $\xi_\mathrm{scat} = 0.0$ it shows that most of the scattered photons are missing at the back of the slab. 
These scattered photons are computed with even small values $\xi_\mathrm{scat}$ as the interaction site for scattering is sampled from the composite function providing reasonable sampling at low {\em and} high optical depths.
At $\xi_\mathrm{scat} = 1.0$, a significant amount of non-Gaussian noise is seen at $\lambda = 151.99~\micron$ due to the extreme amplification of a small number of photons in the calculation.
Setting $\xi_\mathrm{scat} = 0.5$ provides a good compromise between sampling the low probability scattering events and controlling the amplification of the photon weight to be always less than two \citep{Baes16}.

\section{Discussion}

For the $\tauref \leq 1$ cases, the results from the different codes agree within 0.3--2.9\% for the global SEDs and within 3--11\% for the Y slices.
These are near or below the goal precisions for this benchmark.

For the $\tauref = 10$ case, the results from the different codes agree within 1.7--4.0\% for the global SEDs, except for the scattered stellar component where the disagreement is up to 58\%.
The infrared Y slices agree within the goal precisions except for the sto case at 23.10~\micron\ where the disagreement is 20\%.
The optical and UV Y slice deviations are very large, well beyond the goal precisions.
These disagreements in the scattered flux are due to the continued challenge of performing accurate calculations at high optical depths.
The most diagnostic viewing angle for these calculations is $\theta = 180^\circ$ and it is striking that none of the seven codes produces equivalent results with systematic differences up to $10^4$.
While the disagreements are large for the scattered component at high optical depths, the scattered flux at high optical depths is not important as a heating term for the dust emission as evidenced by the agreement between codes for the IR emission SED and images (Fig.~\ref{fig_eff_demis_comp}).
The large differences for the mid-IR wavelengths for DART-Ray and TRADING are due to those codes not calculating the scattered component of the dust emission radiative transfer.
In addition, some of the differences in the IR emission for DART-Ray are due to the lack of inclusion of dust self-heating. 

The lack of agreement or convergence for the scattered component at high optical depths for dust RT codes is shown quantitatively for the first time in this work.  
While it has been known for some time that high optical depths are challenging for dust RT codes, we have shown here that the main issue is with the scattered component and not the directly absorbed component.
Previous benchmarks \citep{Pascucci04, Pinte09} provided tests for high optical depths, but due to the 2D disk geometry used, they do not explicitly test the scattered component at such optical depths.
Further, the 2D disk geometry in previous benchmarks had very high optical depths along the mid-plane of the disk, but very low optical depths along the rotation axis of the system.
Thus, the global scattered flux from these benchmarks is dominated by scattering at low optical depths, not the high disk plane optical depths.
The slab benchmark $\tauref = 10$ case with $\theta = 180^\circ$ explicitly tests the scattered flux at very high optical depths as there are no paths for the scattered photons to the observer that do not go through at least the $\tau_z$ optical depth.
Many of the codes in this benchmark have been run for the previous benchmarks and have given results that are consistent with the literature \citep[e.g.,][]{Bianchi08, Robitaille11, Peest17}.
For the reasons already stated, there is no conflict between reproducing the previous benchmarks at high optical depths and the same codes not agreeing for the scattered component at high optical depths for the benchmark presented in this paper.

For the dust emission approximations, the codes agree best for the effective grain approximation (eff), with somewhat lower precision for the equilibrium-only approximation (equ), and worst for the full solution including stochastic emission (sto).
The larger disagreement for the equ results may be due to the number of dust grain size bins adopted by each model and how the dust properties were averaged or interpolated for these bins.
The larger disagreement for the sto results was expected given the different solution techniques used in the codes for the stochastically heated grains \citep{Camps15b}

The convergence tests provide insight into the potential origin of the differences between codes when combined with Table~\ref{tab_model_params}.
The number of photons each model was run with ($N$ - Table~\ref{tab_model_params}) proved sufficient to reach the desired precision in all cases except for stellar scattering for the $\tauref = 10$. 
Convergence testing indicates that, to reach the desired precision, the number of photon packets would exceed $10^{9}$ with $\tauref = 10$.  
This is a potential cause of some of the large differences between models at UV/optical wavelengths at $\tauref = 10,$ since some models were run with $10^{8}$ photon packets.
But this is not the only cause, as targeted additional tests of just the UV scattering for the $\tauref = 10$ case with many more than $10^9$ photons do not show convergence.
The number of x and y bins ($n_{xy}$) are sufficient in all the model runs.
The number of z bins ($n_z$) are sufficient for all but the DART-Ray results for the $\tauref = 10$ case.
The maximum number of scatterings ($m_\mathrm{scat}$) is of the order of 20 with most codes computing many more scatterings, with the exception of DART-ray and (marginally) SOC.
The maximum number of dust heating iterations ($m_\mathrm{iter}$) is sufficient for all the codes except for DART-Ray for the $\tauref = 10$ case.

The inclusion of six Monte Carlo based radiative transfer codes provided a wealth of comparisons between codes based on the same technique.  
In addition, the inclusion of the DART-Ray code that is based on the alternative technique of Ray-Tracing allowed for comparisons to be made between the two techniques.
This provided ample opportunity to find and remove bugs in all the codes.
The solutions based on Monte Carlo and Ray-Tracing techniques were consistent, with the notable exception of the scattered component in the $\tauref = 10$ case.

\subsection{Lessons learned}

To very carefully setup, specify, and define parameters as much as possible in order to ensure that all the codes perform the same calculation was found to be critical.  
One area that was found to be important was to clearly specify the wavelength grid and ensure that each code performed calculations at the defined wavelengths.
Additionally, the normalization of the slab optical depth was initially at 0.55~\micron, but was changed to be exactly at one of the specified grid points (1~\micron) to avoid,   as
much as possible, interpolation errors in the normalization.
It was also critical to get all the results of the models into the same format and orientations, something that took a surprising amount of time due to the different assumptions made in the different codes.
We also spent significant time establishing a common terminology for different parts of the models and benchmark.

Another lesson learned was that there were minor bugs and/or different conventions that the initial comparisons revealed.
Most of the participating codes made improvements as a result.
These improvements included removing minor bugs that did not significantly change the results but did improve the ability to compare different codes.
Some of these bugs were revealed due to the large parameter space explored by this benchmark, often beyond the range that had been tested in the codes previously.

A major lesson learned was that the codes had significant difficulty with the scattered component for the $\tauref = 10$ case.
This case pushed the codes into an area where dust scattering is critical, both for the stellar and dust emission photons.
The initial results revealed approximately $10^{60}$ differences at the back of the slab in the stellar scattered light for the $\lambda = 0.15~\micron$ Y slice (\S\ref{sec_high_tau_scat}).  
In particular, the Ray-Tracing (DART-ray) results were much higher than many of the Monte Carlo results. 
As Monte Carlo codes have a known limitation in not fully probing the high optical depths, this was not particularly surprising. 
Test cases with some of the Monte Carlo codes at this wavelength with more photons did show smaller differences, but these were still very large. 
Additional test cases were run allowing for larger numbers of scatterings and these also showed smaller differences, but again these were still relatively large.
These differences motivated the inclusion of the composite biasing technique as part of all of the Monte Carlo codes to better probe the high optical depth scatterings \citep{Baes16}.
While the codes still have significant differences for this case and wavelength, they are much smaller at only $\sim 10^4$.
Clearly more work is needed to understand the origin of these differences.
This work has started and has indicated that the origin may be related to very low probability multiple scattering events that dominate the scattered light images at these high optical depths.
While under active investigation, the solution to this issue is beyond the scope of this paper, however this problem did reveal that there may be other very low probability parts of parameter space that are not probed well. 
An example of this is the dust emission when it varies strongly over the model grid.
This is the case for mid-IR wavelengths for $\tauref = 10$ where the dust emission at the front of the slab is many orders of magnitude larger than that from the back of the slab.
Using the dust emission as the probability for emitting a mid-IR photon can lead to very few photons being emitted in the back of the slab.
Of course increasing the number of photons in a run will alleviate this issue, but at the expense of longer run times.
The composite biasing technique can be used to provide for better sampling of the spatial dust emission by, for example, emitting photons one half of the time sampled from the dust emission and one half of the time sampled uniformly in the slab.
This was implemented in the DIRTY code producing much better sampled IR images for the same number of photons emitted.
Another solution may be to use the \citet{Bjorkman01} instantaneous emission technique where the spatial locations of the dust emission photons are based on the locations of dust absorptions, but possibly at the expense of longer run times \citep{Baes05, Chakrabarti09}.

One minor lesson learned was that it would be useful to have the convergence tests run prior to having all the codes run the benchmark.
The convergence tests for this benchmark were generally run after the comparison of results from all the codes was well underway.
While none of the convergence tests revealed that the codes needed to rerun the benchmarks, we did discover that the BASIC wavelength grid for the eff case was close to being too coarse for our goal precision.
The impact of the wavelength grid resolution appeared in the dust emission spectrum with higher resolution grids showing slightly hotter dust.
Convergence tests on the wavelength grid resolution were done with multiple codes for the eff case showing that the global SED convergence was near 1\% for the adopted BASIC grid for the $\tauref = 0.01$ case.
The global SED convergence precision improved for higher \tauref\ values.
While this is within the goal precision, in hindsight we likely would have adopted a finer resolution grid to ensure that this issue was well below the goal precision of 1\%.

We all learned that there is no "easy" benchmark.
The expectation by many of us was that the slab benchmark would be straightforward and take relatively little time to complete.
This was not the case and many of us found this geometrically simple benchmark to be deceptively complex.
This benchmark was useful not only for debugging and refining our codes, but also for deepening our understanding of dust radiative transfer in general.

Many of the lessons learned during the work on this benchmark should be applicable to other 3D dust radiative transfer geometries.
The lack of agreement between the codes for the scattered light at high optical depths ($\tau > 10$) calls for continued caution in interpreting results obtained from any code.
The importance of the convergence tests in illuminating the importance of different parameters in the precision of the solution for this benchmark illustrates that such convergence tests should be done for all geometries.
Convergence tests can be done by anyone using a dust radiative transfer code (not just the coders), and will provide confidence in the results as well as a deeper understanding of the complex dust radiative transfer problem.

\section{Summary}

We present the first 3D dust radiative transfer benchmark.
This benchmark is composed of a rectangular slab of constant density dust externally illuminated by a hot, UV-bright star.
The cases in this benchmark include optical depths from $\tauref = 0.01$--$10$ and three different dust emission assumptions (effective grain, equilibrium-only grains, and the full solution including stochastically heated grains).
Results from seven codes are presented, six based on Monte Carlo techniques and one based on Ray-Tracing.
The results are given as global SEDs and images at selected wavelengths for a range of viewing angles.
The results are in good agreement for $\tauref \leq 1$ with precisions of $\sim$1\% for the global SEDs and $\leq 10\%$ for slices through the images. 
The results are in good agreement for the $\tauref = 10$ case, except for the stellar scattered component.
The setup of this benchmark to explicitly probe the components of the dust radiative transfer problem allowed us to quantify the lack of agreement for the scattered component at high optical depths.
This work provides a benchmark for future 3D RT codes and illustrates remaining challenges for 3D dust RT in the very optically thick regime.

\begin{acknowledgements}
We thank the referee for insightful comments that improved the paper.
This research made use of matplotlib, a Python library for publication quality graphics \citep{Hunter07}.
This research made use of Astropy, a community-developed core Python package for Astronomy \citep{Astropy13}.
RK acknowledges financial support within the Emmy Noether research group on ``Accretion Flows and Feedback in Realistic Models of Massive Star-formation” funded by the German Research Foundation under grant no.~KU 2849/3-1.
MJ acknowledges the support of the Academy of Finland Grant No. 285769.
MB and SB acknowledge support from the European Research Council (ERC) in the form of the FP7 project DustPedia (P.I. Jonathan Davies, proposal 606824).
TL acknowledges the support from the Swedish National Space Board (SNSB) as well as the Chalmers Centre for Computational Science and Engineering (C3SE) and the Swedish National Infrastructure for Computing (SNIC) for providing computational resources. 
G.N. acknowledges support by Leverhulme Trust research project grant RPG-2013-418. 
The development of DART-Ray was supported by the UK Science and Technology Facilities Council (STFC; grant ST/G002681/1).
KG acknowledges the support of Space Telescope Science Institute.
\end{acknowledgements}

%-------------------------------------------------------------------

\bibliographystyle{aa} % style aa.bst
\bibliography{trust_slab}

\end{document}